\documentclass[12pt]{article}
\usepackage{amsmath, amssymb}
\usepackage{tikz}
\usetikzlibrary{arrows.meta, positioning}
\usepackage[round]{natbib}
\usepackage[margin = 2.5cm]{geometry}
\usepackage{url}
\usepackage[colorlinks=true, linkcolor=blue, citecolor=blue, urlcolor=blue]{hyperref}

\newcommand\indep{\protect\mathpalette{\protect\independenT}{\perp}}
\def\independenT#1#2{\mathrel{\rlap{$#1#2$}\mkern2mu{#1#2}}}

\newtheorem{proposition}{Proposition}
\newtheorem{definition}{Definition}
\newtheorem{condition}{Condition}
\newtheorem{remark}{Remark}

\def\Pr{{\mathrm{P}}}
\def\Pdo{{\mathrm{P}^*}}

\DeclareMathOperator{\Anobs}{Anobs}
\DeclareMathOperator{\Paobs}{Paobs}
\DeclareMathOperator{\Chobs}{Chobs}

\tikzset{rv/.style={circle, draw, inner sep=0.5mm, minimum size=7mm}}

\newcommand{\bW}{\boldsymbol{W}}
\newcommand{\bU}{\boldsymbol{U}}
\newcommand{\bV}{\boldsymbol{V}}
\newcommand{\bR}{\boldsymbol{R}}
\newcommand{\bS}{\boldsymbol{S}}
\DeclareMathOperator{\rank}{rank}
\DeclareMathOperator{\Do}{do}
\DeclareMathOperator{\Pa}{Pa}

\def\spacingset#1{\renewcommand{\baselinestretch}%
{#1}\small\normalsize} \spacingset{1}
\newcommand{\blind}{1}

\begin{document}

\if1\blind
{
  \title{\bf Deriving Complete Constraints in Hidden Variable Models}
  \author{Michael C. Sachs$^1$\thanks{EEG and MCS are funded in part by NNF Grant NNF22OC0076595 and the Pioneer Centre for SMARTbiomed}, Erin E. Gabriel$^1$, \\
  Robin J. Evans$^2$,
  Arvid Sj\"olander$^3$\\  
\footnotesize 1:Section of Biostatistics and the Pioneer Centre for SMARTbiomed, \\ \footnotesize University of Copenhagen, Copenhagen, Denmark \\
\footnotesize 2:Department of Statistics, University of Oxford, Oxford, United Kingdom \\
\footnotesize 3:Department of Medical Epidemiology and Biostatistics, Karolinska Institutet, Stockholm, Sweden \\
\footnotesize corresponding author: michael.sachs@sund.ku.dk}
  \maketitle
} \fi 

\if0\blind
{
  \bigskip
  \bigskip
  \bigskip
  \begin{center}
    {\LARGE\bf Deriving Complete Constraints in Hidden Variable Models}
\end{center}
  \medskip
} \fi

\bigskip

\begin{abstract}
Hidden variable graphical models can sometimes imply constraints on the observable distribution that are more complex than simple conditional independence relations. These observable constraints can falsify assumptions of the model and can be used to constrain estimation procedures to improve statistical efficiency. Knowing the complete set of observable constraints is thus ideal, but this can be difficult to determine in many settings.  In models with categorical observed variables and a joint distribution that is completely characterized by linear relations to the probabilities of unobservable response function variables, we develop a systematic method for deriving the complete set of observable constraints. We illustrate the method in several settings, including ones that imply both inequality and equality constraints. 
\end{abstract}

\newpage
\spacingset{1.2}
\vspace{-1cm}

\section{Introduction}

The assumption of a structural model with hidden variables can rule out particular joint distributions not only due to observable conditional independence relations but also due to more complex observable conditions, which may not be clear from inspection of the causal model alone. These observable constraints can be used to falsify a structural model in settings where the data contradict these constraints. The instrumental inequalities first discussed in \citet{pearl1995testability} for the instrumental variable (IV) setting are the most well-known example of these more complex constraints in the form of inequalities, while the ``Verma" constraints \citep{verma2022equivalence} are a well-known example of observable and thus falsifiable equality constraints representing a more general notion of conditional independence. 

In the general IV setting, \citet{bonet2001instrumentality} derived inequality constraints showing that, in cases with an IV with more than two categories, there are observable constraints beyond Pearl's instrumental inequality. Within the specific setting of all discrete observed variables within the classical trivariate IV setting he developed a method for deriving constraints on the observed data distribution implied by the causal assumptions and proved that these are the ``sufficient" set of constraints. We will use the term ``complete'' to refer to what \citeauthor{bonet2001instrumentality} called sufficient, but the meaning remains the same, i.e.~all observable and therefore falsifiable constraints implied by a structural model.  

The \citet{bonet2001instrumentality} method is one example of what \citet{evans2012graphical} called the ``computationally intensive linear programming techniques''. As pointed out by \citet{evans2012graphical}, complete constraints are difficult to derive without using such techniques or algebraic variable elimination algorithms, e.g., Fourier-Motzkin elimination, which become infeasible for moderately sized state spaces. Although other attempts have been made, such as \citet{kang2006inequality}, the derived constraints are not necessarily complete. This is also true for the \emph{nested Markov} characterization which, by construction, only involves equality constraints \citep{evans2019smooth, shpitser2012parameter, verma2022equivalence, richardson2023nested}.

We conceptually extend the method of \citet{bonet2001instrumentality} beyond the IV setting, which due to increased computational power is now feasible in many more settings. We prove that within the class of models that we refer to as ``linear" (cf.~Corollary 2 of \citet{duarte2023automated}), our method provides the complete set of observable constraints. Linear here refers to settings with categorical observed variables in which each district factor is linear in the probabilities of the unobservable response function variables \citep{balke1994counterfactual, sachs2023general}; in our setting, these are the graphs of c-degree 1 (Definition \ref{def:cdeg}). We also relate the equality constraints that we derive to the nested Markov model, and we show that the method can be used outside the ``linear" class to derive at least some of the constraints. We illustrate our method in several examples. All proofs are in the Supplementary Materials, Section \ref{appendix:proofs}, software implementing the method is publicly available on GitHub at \if1\blind{\url{https://github.com/sachsmc/meraconstraints}}\fi \if0\blind{\texttt{[blinded url]}}\fi, and scripts that reproduce the results in the manuscript are provided as ancillary files with this arXiv submission.

\section{Notation and Preliminaries}
\subsection{Notation}
Let $\mathcal{G}(\bV, \bU)$ be a hidden variables directed acyclic graph (DAG) representing a nonparametric structural equations model (NPSEM) \citep{Pearl2000causality}, over variables $\bW = \bV \cup \bU$ where $\bU$ are unobserved and $\bV$ are observed. We assume that all variables in $\bV$ take finitely many values, but no assumptions are made about $\bU$. We will write $\bV_i$ (or some other index) to denote a subset of $\bV$, and $V_i$ a single element of $\bV$. Let $|\bV|$ denote the cardinality of the set $\bV$, and when used for a single variable, for example $|V_i|$, let it denote the number of possible values that variable $V_i$ can take, i.e., the cardinality of its domain. Let $\Paobs(\bV_i)$ denote the union of the sets of observed parents of each variable in the set $\bV_i$ and $\Pa(\bV_i)$ denote the union of the sets of all, observed or unobserved, parents of each variable in the set $\bV_i$. Let $\Anobs(\bV_i)$ denote the union of the sets of observed ancestors of each variable in $\bV_i$, where, as in \citet{evans2019smooth}, $V_2$ is an ancestor of $V_1$ if there is a sequence of directed edges $V_2 \rightarrow \cdots \rightarrow V_1$, and by definition, we include $V_1$ itself in the set of ancestors. Let $\Chobs(\bW_i)$ be the union of the sets of observed children of variables in $\bW_i$, where $W_1$ is a child of $W_2$ if and only if $W_2$ is a parent of $W_1$. Throughout, we assume that the observed distribution is strictly positive, i.e., $\Pr(\bV = v) > 0$ for all $v$, so that all conditional probabilities below are well-defined. The boundary case is discussed in the Supplementary Material, Section \ref{app:positivity}.

\subsection{Settings} \label{sec:settings}
Our goal is to characterize the distribution $\Pr(\bV)$, known as the marginal model (because it is marginal with respect to $\bU$), in terms of constraints implied by the NPSEM for $(\bV,\bU)$. Following \citet{evans2016graphs}, Section 3.2, we can restrict our attention to a smaller class of hidden variables DAGs that satisfy the following two conditions.
\begin{condition} \label{A1}
All unobserved variables are exogenous, i.e., $\Pa(U_i) = \emptyset$ for all $U_i \in \bU$.
\end{condition}

\begin{condition} \label{A2}
The child set of each unobserved variable corresponds to a distinct
set of observed vertices of at least size 2. That is, for each $U_i \in \bU$, $|\Chobs(U_i)| \geq 2$, $\Chobs(U_i) \subseteq \bV$, and there is no other $U_j \in \bU$ such that $\Chobs(U_i) \subseteq \Chobs(U_j)$.
\end{condition}

By Lemmas 1--3 of \citet{evans2016graphs}\footnote{Here we use the references to the published version of these Lemmas, rather than the arXiv version. The arXiv versions are Lemmas 3.7--3.9.}, DAGs where these conditions do not hold can be transformed to DAGs where they do hold without changing the implied marginal model. Briefly, the exogenization process replaces sequences of the form $W_1 \rightarrow U_1 \rightarrow V_1$ by $W_1 \rightarrow V_1$ without changing the marginal model, where $W_1$ could be observed or unobserved. Following exogenization, any unobserved variable $U_1$ whose child set is a subset of that of another unobserved variable $U_2$ can be absorbed into a common unobserved variable with child set equal to the union of child sets of $U_1$ and $U_2$. Examples of these transformations are given in the Supplementary Materials, Section \ref{appendix:trans}.

\begin{definition} \label{def:districts}
Let $\mathcal{G}(\bV, \bU)$ be a hidden variables DAG that by construction obeys Conditions \ref{A1} and \ref{A2}. Remove all edges that do not originate from variables in $\bU$. The sets of variables in $\bW$ that remain connected form a partition of $\bW$ called \emph{districts}. Denote $\mathcal{D}(\mathcal{G}) = \{D_1, \ldots, D_k\}$ the set of districts of $\mathcal{G}$. 
\end{definition}

This same concept is called a \emph{c-component} (short for confounded component) by \citet{tian2002testable}, while the term district is used in, e.g., \citet{richardson2003admgs}, \citet{evans2018margins}, and \citet{duarte2023automated}, which we will follow. We further define the c-degree of a district as follows.

\begin{definition}\label{def:cdeg}
Let $\mathcal{G}(\bV, \bU)$ be a hidden variables DAG that follows Conditions \ref{A1} and \ref{A2}. Let $D_j$ be a district in $\mathcal{G}(\bV, \bU)$. The \emph{c-degree} of $D_j$ is the number of hidden variables $U_i$ in $\bU$ that are in $D_j$. The c-degree of a DAG $\mathcal{G}(\bV, \bU)$ is the maximum of the c-degrees of its districts. 
\end{definition}

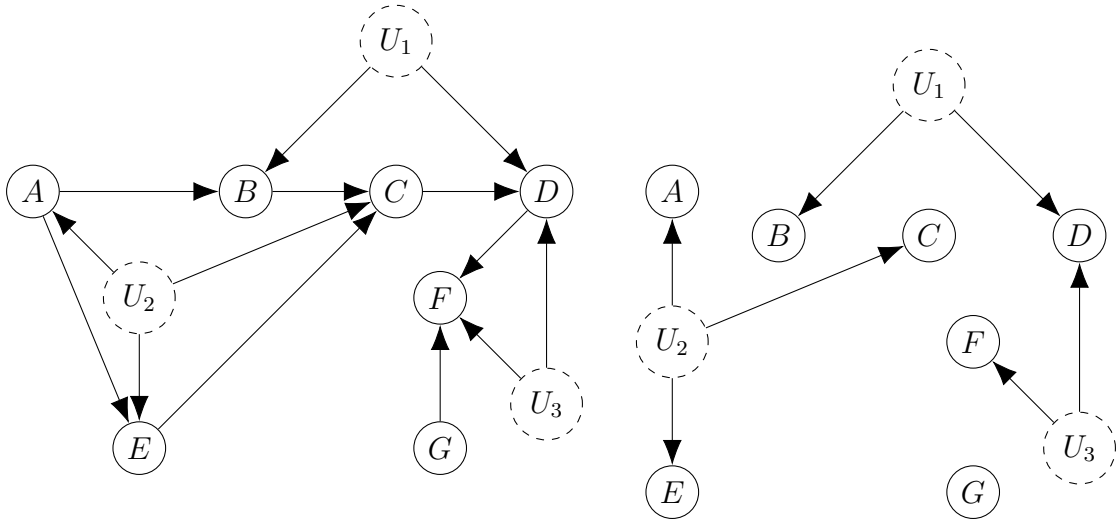
\begin{figure}[ht]
\begin{center}
  \begin{tikzpicture}[>=Stealth, node distance=2cm, on grid, auto]
\begin{scope}
  \node (A) [rv] {$A$};
  \node[draw, circle, dashed] (U2) [below right= of A] {$U_2$};
  \node (B) [above right=of U2, rv] {$B$};
  \node (C) [right=of B, rv] {$C$};
  \node (D) [right=of C, rv] {$D$};
  
  \node (F) [below left=of D, rv] {$F$};
  \node (G) [below= of F, rv] {$G$};

  \node[draw, circle, dashed] (U1) [above =of C] {$U_1$};
  
  \node[draw, circle, dashed] (U3) [below right=of F] {$U_3$};
\node (E) [below=of U2, rv] {$E$};
  \draw [-{Latex[scale=2.0]}]  (A) -- (B);
  \draw [-{Latex[scale=2.0]}]  (E) -- (C);
  \draw [-{Latex[scale=2.0]}]  (C) -- (D);
  \draw [-{Latex[scale=2.0]}]  (D) -- (F);
  \draw [-{Latex[scale=2.0]}]  (A) -- (E);
  \draw [-{Latex[scale=2.0]}]  (B) -- (C);
  \draw [-{Latex[scale=2.0]}]  (G) -- (F);

  \draw [-{Latex[scale=2.0]}] (U1) -- (B);
  \draw [-{Latex[scale=2.0]}] (U1) -- (D);
  \draw [-{Latex[scale=2.0]}] (U2) -- (A);
  \draw [-{Latex[scale=2.0]}] (U2) -- (E);
  \draw [-{Latex[scale=2.0]}] (U2) -- (C);
  \draw [-{Latex[scale=2.0]}] (U3) -- (F);
  \draw [-{Latex[scale=2.0]}] (U3) -- (D);
    
\end{scope}
\begin{scope}[xshift = 8.5cm]
\node (A) [rv] {$A$};
  \node[draw,circle,dashed] (U2) [below= of A] {$U_2$};
  \node (B) [above right=of U2, rv] {$B$};
  \node (C) [right=of B, rv] {$C$};
  \node (D) [right=of C, rv] {$D$};
  
  \node (F) [below left=of D, rv] {$F$};
  \node (G) [below= of F, rv] {$G$};

  \node[draw,circle,dashed] (U1) [above =of C] {$U_1$};
  
  \node[draw,circle,dashed] (U3) [below right=of F] {$U_3$};
\node (E) [below=of U2, rv] {$E$};
  \draw [-{Latex[scale=2.0]}] (U1) -- (B);
  \draw [-{Latex[scale=2.0]}] (U1) -- (D);
  \draw [-{Latex[scale=2.0]}] (U2) -- (A);
  \draw [-{Latex[scale=2.0]}] (U2) -- (E);
  \draw [-{Latex[scale=2.0]}] (U2) -- (C);
  \draw [-{Latex[scale=2.0]}] (U3) -- (F);
  \draw [-{Latex[scale=2.0]}] (U3) -- (D);
      
\end{scope}
\end{tikzpicture}
\end{center}
\caption{Example to illustrate Definitions \ref{def:districts} and \ref{def:cdeg}. The left figure is the original one, and the right figure has the edges originating from observed variables removed. \label{fig:district}}
\end{figure}

The example in Figure \ref{fig:district} illustrates the two definitions in this section. The graph has three districts, one with c-degree 0 ($\{G\}$), one with c-degree 1 ($\{A, C, E, U_2\}$) and one with c-degree 2 ($\{B, D, F, U_1, U_3\}$). Thus, Figure \ref{fig:district} has c-degree 2.

\subsection{Canonical partitioning} \label{sec:respvar}

A DAG that follows Conditions \ref{A1}, \ref{A2}, and that has c-degree 1 comprises districts of either singleton observed variables or sets of observed variables that all share a single common unobserved parent. Such DAGs can be transformed into equivalent mixed graphs (containing directed and undirected edges) that remove unobserved variables $\bU$ having arbitrary distribution and replace them with a set of unobserved variables $\boldsymbol{R}$ that are categorical. These are called \emph{response function variables} by \citet{balke1994counterfactual} and \emph{selectors} by \citet{bonet2001instrumentality}. This is possible because each observed variable is categorical, which induces a canonical partitioning \citep{Pearl2000causality}. See also Theorem 1 of \citet{sachs2023general}. 

\begin{definition}
Let $\mathcal{G}(\bV, \bU)$ be a hidden variables DAG that follows Conditions \ref{A1}, \ref{A2}, and has c-degree 1. The \emph{response function variable transformation} of $\mathcal{G}(\bV, \bU)$ is a mixed graph $\mathcal{G}(\bV, \bR)$ where: 
\begin{itemize}
    \item for each $V_i \in \bV$, introduce a new unobserved variable $R_{V_i}$ that is discrete with $|V_i|^{k_i}$ levels, where $k_i = \prod_{V \in \Paobs(V_i)}|V|$, and add an edge $R_{V_i} \rightarrow V_i$; 
    \item introduce an undirected edge $R_{V_i} \text{---} R_{V_j}$ for every pair of variables $V_i, V_j$ that are in the same district.
\end{itemize}
\end{definition}

\begin{figure}[ht]
\centering
\begin{tikzpicture}[>=Stealth, node distance=2cm, on grid, auto]

  \begin{scope}
  \node (V1) [rv] {$V_1$};
  \node (V2) [below = of V1, rv] {$V_2$};
  \node (V3) [right = of V2, rv] {$V_3$};
  \node (V4) [right = of V3, rv] {$V_4$};
  \node (V5) [below = of V4, rv] {$V_5$};
  \node[draw, circle, dashed] (U2) [below = of V3] {$U_2$};
  \node[draw, circle, dashed] (U3) [below right = of V4] {$U_3$};

  \draw [-{Latex[scale=2.0]}] (V1) -- (V2);
  \draw [-{Latex[scale=2.0]}] (V2) -- (V3);
  \draw [-{Latex[scale=2.0]}] (V3) -- (V4);
  \draw [-{Latex[scale=2.0]}] (V4) -- (V5);
  
  \draw [-{Latex[scale=2.0]}] (U2) -- (V2);
  \draw [-{Latex[scale=2.0]}] (U2) -- (V3);
  \draw [-{Latex[scale=2.0]}] (U3) -- (V4);
  \draw [-{Latex[scale=2.0]}] (U3) -- (V5);
    
  \end{scope}
  \begin{scope}[xshift=8.5cm]  
  \node (V1) [rv] {$V_1$};
  \node (V2) [below = of V1, rv] {$V_2$};
  \node (V3) [right = of V2, rv] {$V_3$};
  \node (V4) [right = of V3, rv] {$V_4$};
  \node (V5) [below = of V4, rv] {$V_5$};
  \node[draw, circle, dashed] (RV1) [right = of V1] {$R_{V_1}$};
  \node[draw, circle, dashed] (RV2) [below = of V2] {$R_{V_2}$};
  \node[draw, circle, dashed] (RV3) [below = of V3] {$R_{V_3}$};
  \node[draw, circle, dashed] (RV4) [right = of V4] {$R_{V_4}$};
  \node[draw, circle, dashed] (RV5) [right = of V5] {$R_{V_5}$};
  
  \draw [-{Latex[scale=2.0]}] (V1) -- (V2);
  \draw [-{Latex[scale=2.0]}] (V2) -- (V3);
  \draw [-{Latex[scale=2.0]}] (V3) -- (V4);
  \draw [-{Latex[scale=2.0]}] (V4) -- (V5);
  
  \draw [-{Latex[scale=2.0]}] (RV1) -- (V1);
  \draw [-{Latex[scale=2.0]}] (RV2) -- (V2);
  \draw [-{Latex[scale=2.0]}] (RV3) -- (V3);
  \draw [-{Latex[scale=2.0]}] (RV4) -- (V4);
  \draw [-{Latex[scale=2.0]}] (RV5) -- (V5);

  \draw[] (RV2) -- (RV3);
  \draw[] (RV4) -- (RV5);
\end{scope}
\end{tikzpicture}
\caption{A sequential instrumental variable example, and its response function variable transformation.\label{fig:seqiv}}
\end{figure}
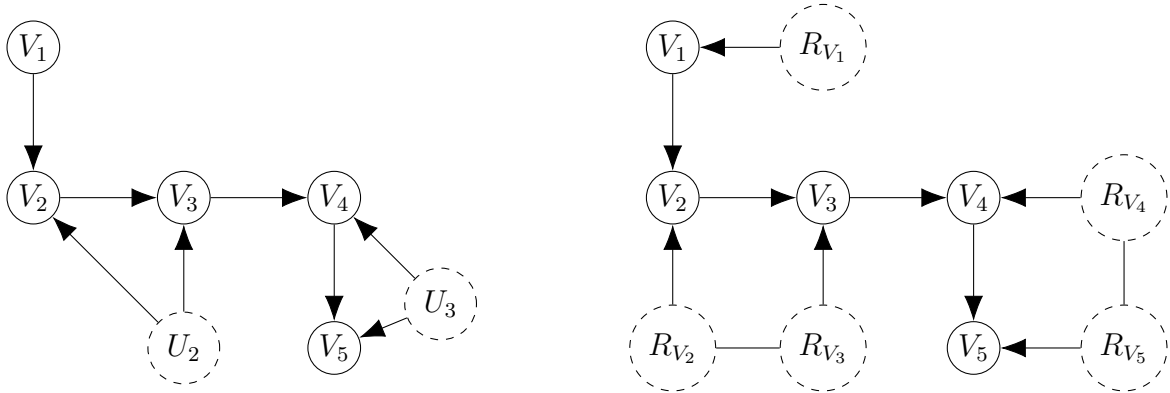

The value of $R_{V_i}$ determines the way that $V_i$ responds to its parents. The undirected edges that connect $R_{V_i}$ and $R_{V_j}$ indicate a statistical dependence, so that $\Pr(R_{V_i} = a, R_{V_j} = b) = \Pr(R_{V_i} = a)\Pr(R_{V_j} = b)$ if $R_{V_i}$ and $R_{V_j}$ are not connected by an undirected edge, but $\Pr(R_{V_i} = a, R_{V_j} = b)$ is not necessarily equal to $\Pr(R_{V_i} = a)\Pr(R_{V_j} = b)$ if they are connected. An example is shown in Figure \ref{fig:seqiv}. We will refer to $R_{V_i}$ as the response function variable of $V_i$, and use the notation $\boldsymbol{R}_{\bV_i}$ to denote the vector of response function variables for the vector of observed variables $\bV_i$. The dependence and independence are joint by construction, not only pairwise. If $\bV^j$ denotes the observed members of the district $D_j$, then the vectors $\boldsymbol{R}_{\bV^1}, \ldots, \boldsymbol{R}_{\bV^k}$ are mutually independent, and the joint distribution of the response function variables within each district is unrestricted apart from the probability simplex constraints.

\begin{remark} \label{rmk1}
Let $\mathcal{G}(\bV, \bU)$ be a hidden variables DAG that follows Conditions \ref{A1}, \ref{A2}, and has c-degree 1. The response function variable transformation of $\mathcal{G}(\bV, \bU)$ results in the same structural model, i.e. the same values of all variables under any intervention (including no intervention) on any of the observed variables in the DAG. 
\end{remark}

The response function variable transformation preserves the marginal model, and also allows us to exploit the discreteness of the observations to formulate a set of linear constraints. Given a value of a response function variable, say $R_{V_i} = b$, there exists a response function $f_{V_i}^b(\Paobs(V_i))$ that determines an observed value for $V_i$ given $\Paobs(V_i)$. Thus $f_{V_i}^b: \operatorname{domain}(\Paobs(V_i)) \rightarrow \operatorname{domain}(V_i)$. Given values of the response function variables for $V_i$ and all of its ancestors, the value of $V_i$ is determined by recursive substitutions \citep{sachs2023general, duarte2023automated}. Therefore we can write all observable probabilities and interventional probabilities in terms of probabilities of response function variables. We will call the equations relating these two forms of probabilities \emph{functional constraints}.

\subsection{Functional constraints}

An algorithm and implementation to obtain the set of functional constraints for DAGs that fit a more strict set of criteria than what we consider here was presented in Algorithm 1 of \citet{sachs2023general}. A more general algorithm for settings with arbitrary c-degree is presented in \citet{duarte2023automated}, where the constraints are polynomials. Their Proposition 4 states that the degree of the polynomial constraints is bounded above by the c-degree of the corresponding district. In the next proposition we explicitly relate the functional constraints to observable probabilities, forming the foundation of the observable constraint derivation.

\begin{proposition} \label{prop:ident}
Suppose $\mathcal{G}(\bV, \bU)$ is a hidden variables DAG that follows Conditions \ref{A1} and \ref{A2}. Further, suppose $\bV_1$ is the vector of all observed variables in the district $D_1$, and let $\bV_2 = \Paobs(\bV_1) \setminus \bV_1$ be the vector of observed parents of $\bV_1$ that are not part of $\bV_1$. We call the elements of $\bV_1$ the \emph{members} of $D_1$ and those of $\bV_2$ its \emph{external parents}. Order $\bV_1=(V_{11},\ldots,V_{1k})$ consistently with a topological ordering of $\mathcal{G}$, let $\bV_1^{<i}=(V_{11},\ldots,V_{1,i-1})$ and $\bV_1^{\leq i} = (V_{11},\ldots,V_{1i})$, and let
\[
\boldsymbol{C}_{i} = \bV_1^{<i} \cup \left\{\bV_2 \cap \Paobs(\bV_1^{\leq i})\right\},
\]
i.e., the earlier members of the district together with the external observed parents of $V_{1i}$ and of every earlier member. Writing $c_i$ for the corresponding subvector of $(v_1, v_2)$, define
\[
    \Pdo(\bV_1=v_1\mid \bV_2=v_2)
    =
    \prod_{i=1}^k
    \Pr\!\left(
        V_{1i}=v_{1i}
        \,\middle|\,
        \boldsymbol{C}_{i} = c_i
    \right).
\]
If $D_1$ has c-degree at most 1, then for all possible values of the vectors $v_1 = (v_{11}, \ldots, v_{1k}), v_2 = (v_{21}, \ldots, v_{2\ell})$, the functional constraints on $\Pdo(\bV_1 = v_{1} \mid \bV_2 = v_2)$ under $\mathcal{G}(\bV, \bU)$ are linear functions of response function variable probabilities of $\bR_{\bV_1}$.
\end{proposition}

We call $\Pdo$ the \emph{district factor} of $D_1$. Under the conditional independencies implied by $\mathcal{G}$, conditioning additionally on any other variables that precede $V_{1i}$ in a topological ordering does not change the factor (see the proof of Proposition \ref{prop:ident}), but $\boldsymbol{C}_i$ cannot be made smaller in general.

Let $\Pr(\bV_1 = v_1 \mid \Do(\bV_2 = v_2))$ denote the joint probability of $\bV_1 = v_1$ under the intervention that sets $\bV_2 = v_2$. 

\begin{proposition} \label{causalprop}
Suppose $\mathcal{G}(\bV, \bU)$ is a hidden variables DAG that follows Conditions \ref{A1}, \ref{A2}, and $\bV_1$, $\bV_2$ are as described in Proposition \ref{prop:ident}. Then all $\Pr(\bV_1 = v_{1} \mid \Do(\bV_2 = v_2))$ are identified and given by $\Pdo(\bV_1 = v_1 \mid \bV_2=v_2)$. 
\end{proposition}

\begin{figure}[ht]
\centering

    \begin{tikzpicture}
    \begin{scope}[xshift=-2cm]
    \node[draw, circle] at (2, 0) (X) {\(X\)};
    \node[draw, circle] at (4, 0) (Y) {\(Y\)};
    \node[draw, circle] at (0, 0) (Z) {$Z$};
    \node[draw, circle, dashed] at (3,2) (U) {\(U\)};
    \draw [-{Latex[scale=2.0]}] (X) to (Y);
    \draw [-{Latex[scale=2.0]}] (Z) to (X);
    \draw [-{Latex[scale=2.0]}] (U) to (X);
    \draw [-{Latex[scale=2.0]}] (U) to (Y);
    \draw[dashed] (4.75, -1) -- (4.75, 3);

    \end{scope}
\begin{scope}[xshift=3.5cm]
    \node[draw, circle] at (2, 0) (X) {\(X\)};
    \node[draw, circle] at (4, 0) (Y) {\(Y\)};
    \node[draw, circle] at (0, 0) (Z) {$Z$};
    \node[draw, circle, dashed] at (3,2) (U) {\(U\)};
    \draw [-{Latex[scale=2.0]}] (U) to (X);
    \draw [-{Latex[scale=2.0]}] (U) to (Y);
    \draw[dashed] (4.75, -1) -- (4.75, 3);

    \end{scope}
    
  \begin{scope}[xshift=9cm] 
    
    \node[draw, circle] at (2, 0) (X) {\(X\)};
    \node[draw, circle] at (4, 0) (Y) {\(Y\)};
    \node[draw, circle] at (0, 0) (Z) {$Z$};
    \node[draw, circle, dashed] at (2,2) (Rx) {\(R_x\)};
    \node[draw, circle, dashed] at (4,2) (Ry) {\(R_y\)};
    \node[draw, circle, dashed] at (0,2) (Rz) {\(R_z\)};
    \draw [-{Latex[scale=2.0]}] (X) to (Y);
    \draw [-{Latex[scale=2.0]}] (Z) to (X);
    \draw [-{Latex[scale=2.0]}] (Rx) to (X);
    \draw [-{Latex[scale=2.0]}] (Ry) to (Y);
    \draw [-{Latex[scale=2.0]}] (Rz) to (Z);
    \draw (Rx) -- (Ry);

  \end{scope}
    \end{tikzpicture}
  
    \caption{\label{fig:ivclassic} The instrumental variable DAG in original form (left), showing the two districts (center), and response function variable form (right).}
    \end{figure}
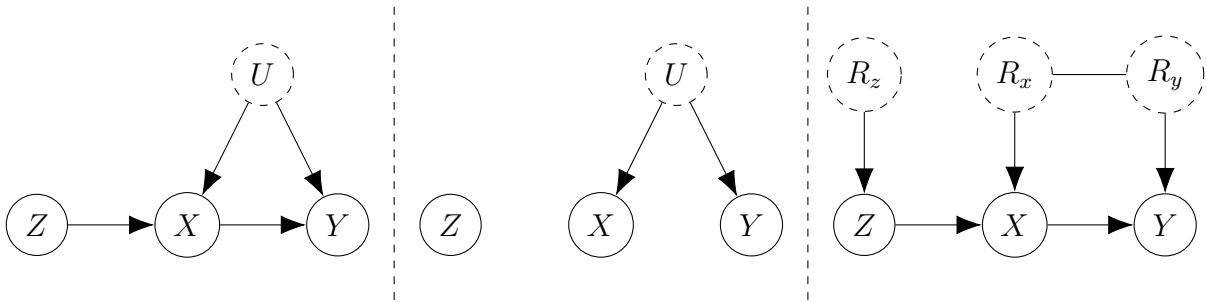

Consider the example in Figure \ref{fig:ivclassic}, the leftmost panel of which is the classical IV setting. In this case there are 2 districts, $\{Z\}$ with c-degree 0 and $\{X, Y, U\}$ with c-degree 1, and Conditions \ref{A1} and \ref{A2} are satisfied. Letting $\bV_1 = (X, Y)$ and applying Proposition \ref{prop:ident} we have $\bV_2 = \{Z\}$, $\boldsymbol{C}_1 = \{Z\}$, $\boldsymbol{C}_2 = \{Z, X\}$, and
\begin{align*}
    \Pdo(\bV_1 = v_{1} \mid \bV_2 = v_2) &= \Pr(Y = y \mid X = x, Z = z) \Pr(X = x \mid Z = z)\\ 
    & = \Pr(X = x, Y = y \mid \Do(Z = z)) \\
    & = \sum_{b \in \mathcal{B}} \Pr((R_x, R_y) = b), \mbox{ for all } x, y, z,
\end{align*}
where $\mathcal{B} = \{(b_x, b_y): f^{b_y}_Y(f^{b_x}_X(z)) = y \mbox{ and } f^{b_x}_X(z) = x\}$ for the response functions $f^{b_y}_Y$ and $f^{b_x}_X$. The series of functional constraints for each combination of values of $z, x, y$ can be represented by the matrix equation
\[
\boldsymbol{p_{xy\cdot z}} = \boldsymbol{B} \boldsymbol{r_{xy}}, 
\]
where, for binary $X, Y, Z$, $\boldsymbol{p_{xy\cdot z}}$ is the vector of 8 probabilities $[\Pr(Y = y \mid X = x, Z = z) \Pr(X = x \mid Z = z): x, y, z \in \{0, 1\}]$, $\boldsymbol{B}$ is an $8 \times 16$ matrix of 0s and 1s, and $\boldsymbol{r_{xy}}$ is the vector of 16 response function variable probabilities $[\Pr((R_x, R_y) = (b_x, b_y)): b_x, b_y \in \{0, 1, 2, 3\}]$.

\subsection{Nested Markov models} \label{sec:nested}

The nested Markov model \citep{richardson2023nested} constrains the observed distribution through (generalized) conditional independence relations. Briefly put, it allows these relations to be encoded within a hidden variables DAG model, and in the discrete case they are the only \emph{equality} constraints in the model \citep{evans2018margins}.

While a regular conditional independence relation, for example $X \indep Y \mid Z$, can be expressed as $\Pr(X \mid Y, Z) = \Pr(X \mid Z)$, a \emph{nested Markov independence} is a more general statement about a lack of dependence of a quantity on a variable. In particular, the quantities in question are probabilities in graphs after removing arrows into certain variables (i.e., intervening on them). For example, they can be expressed as $\sum_B \Pr(D \mid A, B, C)\Pr(B \mid A) = g(D, C)$, for some function $g$; note the lack of dependence on $A$ here. Such relations are also sometimes called \emph{Verma constraints} after \citet{verma2022equivalence}, and \emph{post-truncation independences} after \citet{shpitser2012parameter}.
Further details are given in the Supplementary Materials, Section \ref{sec:nmm}.

\section{Method for deriving constraints and completeness results}

Given those preliminaries, we now detail our algorithm for deriving observable constraints in a hidden variables DAG with discrete and finite observed variables. To summarize briefly, we begin by using d-separation to list all conditional and marginal independencies implied by the DAG, if there are any. Then, for each district, we describe a convex polyhedron in terms of the probabilities of the response function variables and convert this to an equivalent description in terms of only observable quantities. This is called the vertex-to-halfspace (V-to-H) conversion, and different algorithms exist to efficiently perform this conversion that are implemented in open-source software programs such as \texttt{cddlib} \citep{cdd}.

\subsection{Description of the algorithm} \label{algorithm}

\textbf{Inputs: } A hidden variables DAG $\mathcal{G}(\bV, \bU)$ that meets Conditions \ref{A1}, \ref{A2} and has c-degree 1. 
\begin{enumerate}
    \item Enumerate the conditional independencies using a standard algorithm based on checking for d-separation, e.g., \citet{van2014constructing}, or using lines 1--8 of Algorithm 1 of \citet{richardson2023nested}.
    \item[] \, \textbf{Output: } List of standard independence constraints.  
    \item Identify the districts $\mathcal{D} = \{D_1, \ldots, D_M\}$. 
    \item Do the response function variable transformation of $\mathcal{G}$ into $\mathcal{G}(\bV, \bR)$ as described in Section \ref{sec:respvar}.
    \item For each district $D_j$: 
   \begin{enumerate}  
        \item Derive the formula for $\Pdo(\bV_1 = v_1 \mid \bV_2 = v_2)$, where $\bV_1$ are all the observed variables in $D_j$ and $\bV_2 = \Paobs(\bV_1) \setminus \bV_1$, using Proposition \ref{prop:ident}.
        \item For each combination of $v_1, v_2$, derive the functional probabilities, i.e., 
        \[
        \sum_{b \in \mathcal{B}} \Pr(\boldsymbol{R}_{\bV_1} = b), 
        \]
    where $\mathcal{B}$ is the set of response function variable values that are compatible with the observation vector $v_1, v_2$. 
    \item Construct the matrix representing the system of linear equations relating 
    \[
    \Pdo(\bV_1 = v_1 \mid \bV_2 = v_2) = \sum_{b \in \mathcal{B}} \Pr(\boldsymbol{R}_{\bV_1} = b) \Leftrightarrow \boldsymbol{p} = \boldsymbol{B} \boldsymbol{r}.
    \]
    \item Add a row of all 1s to $\boldsymbol{B}$ and a 1 to $\boldsymbol{p}$ which represents the constraint that the response function variable probabilities all sum to 1. 
    \item Convert $\boldsymbol{p} = \boldsymbol{B} \boldsymbol{r}$ to the H-representation, e.g., using the double description method \citep{cdd, doubledescriptionmethod}.
    \item[] \, \textbf{Output: } The H-representation and the vector $\boldsymbol{p}$ of the formulae for $\Pdo(\bV_1 = v_1 \mid \bV_2 = v_2)$ for each district
    \end{enumerate}
\end{enumerate}

The H-representation consists of two rational matrices $H_i, H_e$ and rational vectors $b_i, b_e$ such that 
\[
H_i \boldsymbol{p} \leq b_i
\mbox{ and }
H_e \boldsymbol{p} = b_e. 
\]
In other words, it provides a representation of inequality and equality constraints in terms of the observable probability vector.

\subsection{Completeness results} \label{sec:complete}

\begin{proposition} \label{prop:complete}
 Let $\mathcal{G}$ be a hidden variables DAG satisfying Conditions \ref{A1}, \ref{A2}, and having c-degree 1. Let $\mathcal{C}(\mathcal{G})$ be the set of strictly positive observed distributions satisfying the conditional independencies implied by $\mathcal{G}$ and the district level H-representation constraints returned by the algorithm. Then
\[
    \mathcal{M}(\mathcal{G}) = \mathcal{C}(\mathcal{G}),
\]
where $\mathcal{M}(\mathcal{G})$ denotes the strictly positive distributions in the marginal model over the observed variables. Hence the constraints derived from the algorithm, together with the conditional independencies, are complete.
\end{proposition}

In other words, if every district in a DAG has at most one unobserved common cause, then the constraints derived by our algorithm are necessary and sufficient, i.e., the marginal model is exactly the set of observed distributions satisfying these constraints.

As a case where the algorithm does not apply, consider the triangle scenario, in which three observed variables $A$, $B$, $C$ have no edges between them and each pair shares an unobserved parent. It has c-degree 3, and comprises a single district. Our algorithm cannot be applied in this case; however, it is known that some observed probability distributions are incompatible with the triangle scenario. For example, \citet{Wolfe_2019} demonstrates that the distribution where $\Pr(A = 1, B = 1, C = 1) = \Pr(A = 0, B = 0, C = 0) = 1/2$ is incompatible with the triangle scenario. An additional example is the left side of Figure \ref{fig:incomp}. As discussed in Section \ref{sec:nonlin}, one can sometimes modify graphs with c degree greater than 1 to obtain some or all constraints. 

We note that our restriction to strictly positive observed distributions ensures that all probabilities in the district factors are well-defined. Once the symbolic constraints are derived, they can of course be applied to estimated distributions that include zero probabilities, as that does not necessarily imply that the true distribution has zero probabilities. We consider that case where it is of interest to derive constraints under additional structural assumptions that impose events of probability zero in the Supplementary Materials, Section \ref{app:positivity}

\subsection{Identifying nontrivial constraints} \label{sec:nontriv}

Not every constraint returned by the algorithm restricts the observed distribution. The equality constraints include normalization, as in \eqref{eqn:ivex}, and the inequality constraints include nonnegativity, $\Pdo(\bV_1 = v_1 \mid \bV_2 = v_2) \geq 0$. Other equality constraints follow from the conditional independencies, which are listed separately in step 1 of the algorithm, and other inequality constraints become nonnegativity constraints once equality constraints are added to them. Throughout this subsection, let $D$ be a district with members $\bV_1$ and external parents $\bV_2$, write $p(v_1 \mid v_2) = \Pdo(\bV_1 = v_1 \mid \bV_2 = v_2)$ for the entries of the vector $\boldsymbol{p}$, and let $\boldsymbol{\check{P}}$ be the cross-product of probability simplices, one for each value of $v_2$. This is the set of vectors allowed by the standard probabilistic constraints alone.

\begin{definition} \label{def:nontrivial}
An equality constraint of $D$ is \emph{trivial} if it holds for every strictly positive distribution of the observed variables that satisfies the conditional independencies implied by $\mathcal{G}$. An inequality constraint of $D$ is \emph{trivial} if it holds for every $\boldsymbol{p} \in \boldsymbol{\check{P}}$ that satisfies the equality constraints of $D$. A constraint that is not trivial is \emph{nontrivial}.
\end{definition}

The trivial equality constraints can be listed from the graph. Call a set $\bS \subseteq \bV_1$ \emph{closed} if it contains every member of $D$ that is an ancestor of a member of $\bS$. Equivalently, $\bS$ consists of the first members in some ordering of $\bV_1$ that is consistent with a topological ordering of $\mathcal{G}$. Let $\boldsymbol{E}(\bS) = \bV_2 \cap \Paobs(\bS)$ be the external parents of $\bS$. Consider the linear equations
\begin{itemize}
\item[(N)] $\sum_{v_1} p(v_1 \mid v_2) = 1$ for every $v_2$;
\item[(O)] for every nonempty closed set $\bS$, the margin $p_{\bS}(v_{\bS} \mid v_2) = \sum_{v_{\bV_1 \setminus \bS}} p(v_1 \mid v_2)$ takes the same value at any two values of $v_2$ that agree on $\boldsymbol{E}(\bS)$.
\end{itemize}
By Proposition \ref{causalprop}, $p_{\bS}(v_{\bS} \mid v_2)$ is the probability of $\bS = v_{\bS}$ under $\Do(\bV_2 = v_2)$, and (O) states that it does not depend on the external parents of the members outside $\bS$. We also write $N$ and $O$ for the coefficient matrices of these equations.

\begin{proposition} \label{prop:nontrivial}
Suppose $\mathcal{G}(\bV, \bU)$ is a hidden variables DAG that follows Conditions \ref{A1}, \ref{A2} and has c-degree 1. Let $H_i \boldsymbol{p} \leq b_i$, $H_e \boldsymbol{p} = b_e$ be the H-representation of a district $D$ returned by the algorithm, in which each row of $H_i$ defines a different facet of the district polytope $\{\boldsymbol{B}\boldsymbol{r} : \boldsymbol{r} \text{ a probability vector}\}$ and the rows of $H_e$ define its affine hull. Then:
\begin{enumerate}
\item[(i)] An equality constraint $g^\top \boldsymbol{p} = d$ is trivial if and only if it is a linear combination of the equations (N) and (O). Hence the number of linearly independent nontrivial equality constraints of $D$ is
\[
\rank\begin{pmatrix} N \\ O \\ H_e \end{pmatrix} - \rank\begin{pmatrix} N \\ O \end{pmatrix},
\]
and any rows of $H_e$ that extend a basis of the rows of $(N; O)$ to a basis of the rows of $(N; O; H_e)$ imply, together with the conditional independencies, all equality constraints of $D$.
\item[(ii)] An inequality constraint $h^\top \boldsymbol{p} \leq c$ in $H_i$ is trivial if and only if some entry $p_k$ of $\boldsymbol{p}$ is zero at every column $\beta$ of $\boldsymbol{B}$ with $h^\top \beta = c$. In that case $c - h^\top \boldsymbol{p} = \alpha p_k + \lambda^\top (H_e \boldsymbol{p} - b_e)$ for some $\alpha > 0$ and vector $\lambda$, so the constraint is the nonnegativity constraint $p_k \geq 0$ written in a different form. If the constraint is nontrivial, then there is a strictly positive distribution of the observed variables that satisfies the conditional independencies and every constraint of every district except this one, which it violates.
\end{enumerate}
\end{proposition}

By Proposition \ref{prop:complete}, for strictly positive distributions, removing the trivial constraints from the output of the algorithm does not change the model, and by part (ii), removing a nontrivial inequality constraint enlarges it. The equations (O) must be imposed for every closed set, not only for the sets $\bV_1^{\leq i}$ of Proposition \ref{prop:ident}. For the sets $\bV_1^{\leq i}$, (O) holds for every strictly positive distribution, because $p_{\bV_1^{\leq i}}$ is the product of the first $i$ factors in Proposition \ref{prop:ident}. For the other closed sets, it is a consequence of the conditional independencies.

Neither part of Proposition \ref{prop:nontrivial} requires optimization. Part (i) uses rank computations. For part (ii), let $I$ be the 0/1 matrix, computed in exact rational arithmetic, with $I_{r\beta} = 1$ when column $\beta$ of $\boldsymbol{B}$ lies on the facet defined by row $r$ of $H_i$. Then row $r$ is nontrivial if and only if every entry of row $r$ of $I \boldsymbol{B}^\top$ is positive.

The equations (O) are the subset of nested Markov constraints that are ordinary independencies. We can distinguish them from the generalized independencies as follows. Call a set $\bS \subseteq \bV_1$ \emph{closed in the district} if it contains every member of $D$ that is a parent of a member of $\bS$. For such a set, the members of $\bS$ are functions of $v_{\boldsymbol{E}(\bS)}$ at every column of $\boldsymbol{B}$, so $p_{\bS}$ depends on $v_2$ only through its value on $\boldsymbol{E}(\bS)$, and this is an equality constraint of $D$. These are the constraints that the nested Markov model places on the district factor: $p_{\bS}$ is the kernel obtained by fixing the members outside $\bS$, and the constraint states that it does not depend on $\bV_2 \setminus \boldsymbol{E}(\bS)$ \citep{richardson2023nested}. When $\bS$ is closed, the constraint is of type (O), and by part (i) it is implied by the ordinary conditional independencies. A set can be closed in the district but not closed, because a member can be an ancestor of another member through variables outside $D$, and then the constraint is in general a Verma constraint. In the graph $V_1 \rightarrow V_2 \rightarrow V_3 \rightarrow V_4$ with $U \rightarrow V_2$ and $U \rightarrow V_4$, for example, the district $\{V_2, V_4\}$ has external parents $\{V_1, V_3\}$, and the set $\{V_4\}$ is closed in the district but not closed. Its margin is
\[
\sum_{v_2} \Pdo(V_2 = v_2, V_4 = v_4 \mid V_1 = v_1, V_3 = v_3) = \sum_{v_2} \Pr(v_2 \mid v_1)\Pr(v_4 \mid v_1, v_2, v_3),
\]
and the constraint states that it does not depend on $v_1$, as in the example of Section \ref{sec:nested}. With binary variables, these are the 2 nontrivial equality constraints that part (i) finds in this district. Since the nested Markov constraints are the only equality constraints of the marginal model \citep{evans2018margins}, every nontrivial equality constraint is a Verma constraint, and in all examples in this paper, the margin equations of the sets that are closed in the district but not closed span them.

Part (ii) must take the equality constraints into account. A row of $H_i$ is determined only up to a positive multiple and the addition of a linear combination of the rows of $H_e$, and which representative the double description method returns depends on its input. For example, in the graph $X \rightarrow A \leftarrow U \rightarrow B \leftarrow Y$ with binary variables, writing $p(ab \mid xy)$ for $\Pdo(A = a, B = b \mid X = x, Y = y)$, one of the returned rows is
\[
p(00 \mid 11) \leq p(00 \mid 10) + p(01 \mid 10).
\]
It is violated at points of $\boldsymbol{\check{P}}$ with $p(00 \mid 11) = 1$ and $p(00 \mid 10) = p(01 \mid 10) = 0$. However, adding the equality constraint $p(00 \mid 10) + p(01 \mid 10) = p(00 \mid 11) + p(01 \mid 11)$, which is of type (O) with $\bS = \{A\}$, turns it into $p(01 \mid 11) \geq 0$. The 12 columns of $\boldsymbol{B}$ on its facet all have $p(01 \mid 11) = 0$. Of the 24 inequality constraints of this district, Proposition \ref{prop:nontrivial} identifies 8 as nontrivial, the CHSH inequalities \citep{clauser1969proposed}, and none of its 8 equality constraints is nontrivial.

We apply the algorithm to the second district of the binary IV example which contains $X$ and $Y$. The worked details are in the Supplementary Materials, Section \ref{appendix:iv}. The equality constraints encoded by $H_e$ are simply the standard probabilistic constraints:
\begin{equation}
\sum_{x, y \in \{0, 1\}} \Pr(Y = y, X = x \mid Z = z) = 1, \mbox{ for } z = 0, 1. \label{eqn:ivex}
\end{equation}
The closed sets are $\{X\}$ and $\{X, Y\}$. Both have external parents $\{Z\}$, so (O) has no equations, and the two equality constraints \eqref{eqn:ivex} are the normalization constraints (N), which are trivial. Of the 12 inequality constraints, 8 are nonnegativity constraints. Proposition \ref{prop:nontrivial} identifies the other 4, the 2nd, 3rd, 5th and 6th rows of $H_i$ in the Supplementary Materials, Section \ref{appendix:iv}, as nontrivial. These are (after some simple algebra) the classical instrumental inequalities of \citet{pearl1995testability}:
\begin{align*}
 & \Pr(Y = 1,X = 1 \mid Z = 0) + \Pr(Y = 0,X = 1 \mid Z = 1) \leq 1 \\
& \Pr(Y = 0,X = 1 \mid Z = 0) + \Pr(Y = 1,X = 1 \mid Z = 1)  \leq 1 \\
& \Pr(Y = 1,X = 0 \mid Z = 0) + \Pr(Y = 0,X = 0 \mid Z = 1) \leq 1 \\
& \Pr(Y = 0,X = 0 \mid Z = 0) + \Pr(Y = 1,X = 0 \mid Z = 1) \leq 1.
\end{align*}
Singleton districts with no parents, such as $\{Z\}$, have only the trivial constraints (N) and nonnegativity. The sequential instrumental variable model of Figure \ref{fig:seqiv} has the conditional independence $(V_1, V_2) \indep (V_4, V_5) \mid V_3$ and two sets of instrumental inequalities, one for each of the districts $\{V_2, V_3\}$ and $\{V_4, V_5\}$.

The front door model $A \rightarrow M \rightarrow Y$ with $U \rightarrow A$ and $U \rightarrow Y$ is known to have no nontrivial constraints \citep{bhattacharya2022testability}. Its district $\{A, Y\}$ has external parent $M$. The closed sets are $\{A\}$, because $A$ is an ancestor of $Y$ through $M$, and $\{A, Y\}$. For $\bS = \{A\}$, (O) states that $\sum_y \Pdo(A = a, Y = y \mid M = m)$ does not depend on $m$, which in observed probabilities reads $\Pr(A = a) = \Pr(A = a)$. Together with normalization, these equations account for all 3 equality constraints of the district, so none of them is nontrivial. Each of the 8 inequality constraints is a nonnegativity constraint, so Proposition \ref{prop:nontrivial} correctly finds no nontrivial constraints.

\section{Examples}
\subsection{Nested equivalent models with distinct inequalities}

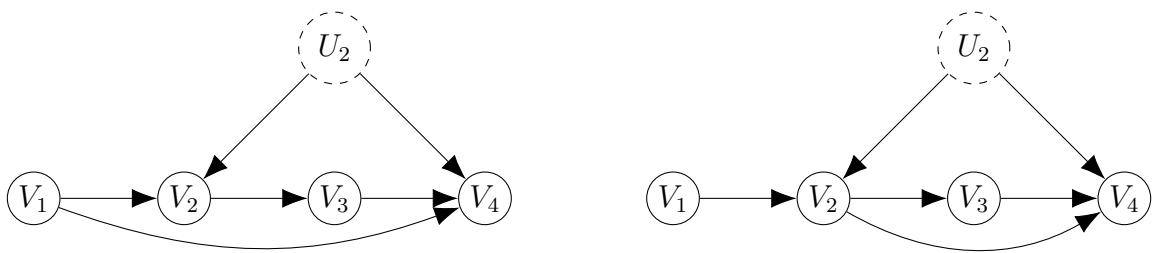
\begin{figure}[ht]
    \centering
\begin{tikzpicture}[>=Stealth, node distance=2cm, on grid, auto]
  \begin{scope}
  \node (V1) [rv] {$V_1$};
  \node (V2) [right = of V1, rv] {$V_2$};
  \node (V3) [right = of V2, rv] {$V_3$};
  \node (V4) [right = of V3, rv] {$V_4$};
  \node[draw,circle,dashed] (U2) [above = of V3] {$U_2$};

  \draw [-{Latex[scale=2.0]}] (V1) -- (V2);
  \draw [-{Latex[scale=2.0]}] (V2) -- (V3);
  \draw [-{Latex[scale=2.0]}] (V3) -- (V4);
  \draw [-{Latex[scale=2.0]}] (V1) to[bend right=20] (V4);
  
  \draw [-{Latex[scale=2.0]}] (U2) -- (V4);
  \draw [-{Latex[scale=2.0]}] (U2) -- (V2);
\end{scope}

    \begin{scope}[xshift=8.5cm]
  \node (V1) [rv] {$V_1$};
  \node (V2) [right = of V1, rv] {$V_2$};
  \node (V3) [right = of V2, rv] {$V_3$};
  \node (V4) [right = of V3, rv] {$V_4$};
  \node[draw,circle,dashed] (U2) [above = of V3] {$U_2$};

  \draw [-{Latex[scale=2.0]}] (V1) -- (V2);
  \draw [-{Latex[scale=2.0]}] (V2) -- (V3);
  \draw [-{Latex[scale=2.0]}] (V3) -- (V4);
  \draw [-{Latex[scale=2.0]}] (V2) to[bend right] (V4);

  \draw [-{Latex[scale=2.0]}] (U2) -- (V4);
  \draw [-{Latex[scale=2.0]}] (U2) -- (V2);
\end{scope}
\end{tikzpicture}
    \caption{Two graphs that are nested Markov equivalent, but which have distinct inequality constraints.}
    \label{fig:nested-ineq}
\end{figure}

Figure \ref{fig:nested-ineq} shows two graphs that have the same equality constraints, and hence the same nested Markov model, but different inequality constraints. Both graphs satisfy Conditions \ref{A1} and \ref{A2}, have c-degree 1, and have the districts $\{V_1\}$, $\{V_2, V_4, U_2\}$ and $\{V_3\}$, of which $\{V_1\}$ and $\{V_3\}$ have only trivial constraints. Both graphs imply $V_1 \indep V_3 \mid V_2$, and in both the district $\{V_2, V_4, U_2\}$ has the factor
\begin{align*}
& \Pdo(V_2 = v_2, V_4 = v_4 \mid V_1 = v_1, V_3 = v_3) \\
& \quad = \Pr(V_2 = v_2 \mid V_1 = v_1) \Pr(V_4 = v_4 \mid V_2 = v_2, V_1 = v_1, V_3 = v_3).
\end{align*}
Its 6 equality constraints are the same in both graphs. By Proposition \ref{prop:nontrivial}(i) they are all trivial: they are normalization and (O) with $\bS = \{V_2\}$, which this factorization satisfies identically. Thus the conditional independence is the only equality constraint of either graph.

The 16 inequality constraints of the left graph are all nonnegativity constraints, so it has no other restrictions. In the right graph, Proposition \ref{prop:nontrivial}(ii) identifies 8 of the 24 inequality constraints as nontrivial, and the other 16 are nonnegativity constraints. With binary variables, and after adding equality constraints to some of the rows returned by the algorithm, the 8 nontrivial constraints are
\begin{equation}
\Pdo(V_2 = v_2, V_4 = v_4 \mid V_1 = 0, V_3 = v_3) + \Pdo(V_2 = v_2, V_4 = 1 - v_4 \mid V_1 = 1, V_3 = v_3) \leq 1, \label{eqn:nestedineq}
\end{equation}
for $v_2, v_3, v_4 \in \{0, 1\}$. For each value of $v_3$, these are the instrumental inequalities of \citet{pearl1995testability}, with instrument $V_1$, treatment $V_2$ and outcome $V_4$.

The inequalities \eqref{eqn:nestedineq} express that $V_1$ affects $V_4$ only through $V_2$ and $V_3$. Write $V_4(v_2, v_3)$ for the value of $V_4$ under $\Do(V_2 = v_2, V_3 = v_3)$, and $V_2(v_1)$ similarly. In the right graph, $V_4(v_2, v_3)$ does not depend on $v_1$, so the left side of \eqref{eqn:nestedineq} is
\[
\Pr(V_2(0) = v_2, V_4(v_2, v_3) = v_4) + \Pr(V_2(1) = v_2, V_4(v_2, v_3) = 1 - v_4).
\]
These two events are mutually exclusive, because they require the same potential outcome to take different values, so the sum is at most 1. A violation rules out the right graph. We would expect one if $V_1$ changes $V_4$ with $V_2$ and $V_3$ held fixed, as the edge $V_1 \rightarrow V_4$ in the left graph allows. The left graph also has the exclusion restriction that $V_2$ affects $V_4$ only through $V_3$, but this restriction has no observable consequences beyond the conditional independence: given $V_1$, the unobserved $U_2$ can determine $V_2$, so $V_4$ can depend on $V_2$ through $V_1$ and $U_2$.

\subsection{Tripartite Bell graphs} \label{sec:tripart}

\begin{figure}[ht]
\begin{center}
\begin{tikzpicture}[
    >=Latex,
    var/.style   ={draw,circle,minimum size=9mm,inner sep=0pt},
    latent/.style={draw,dashed,circle,minimum size=9mm,inner sep=0pt}
]
\node[latent]  (L) at (0,0)    {$U$};

\node[rv] (A) at ( 2.2, 0.8) {$V_1$};
\node[rv] (X) at ( 2.2,-1.2) {$X$};

\node[rv] (C) at ( 0.0,-2.0) {$V_3$};
\node[rv] (Z) at (-2.2,-1.2) {$Z$};

\node[rv] (B) at (-2.2, 0.8) {$V_2$};
\node[rv] (Y) at ( 0.0, 2.0) {$Y$};

\draw[->] (X) -- (A);
\draw[->] (Y) -- (B);
\draw[->] (Z) -- (C);

\draw[->] (L) -- (A);
\draw[->] (L) -- (B);
\draw[->] (L) -- (C);

\end{tikzpicture}
\end{center}
\caption{The tripartite Bell scenario. \label{fig:tripart}}
\end{figure}
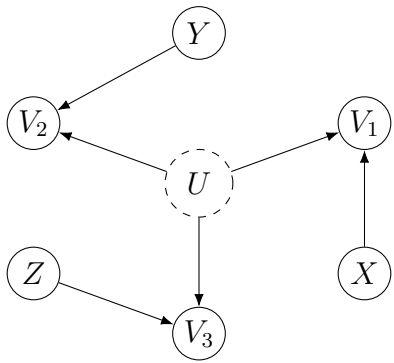

Figure \ref{fig:tripart} is a well-known causal model in quantum physics. An incomplete set of constraints has been derived using the entropy method in \citet{bell3}. Our algorithm finds the complete set of constraints in this setting in about 19 minutes (Table \ref{tab:compute}). First we identify the standard independencies $X \indep (V_2, V_3, Y, Z)$, $Y \indep (V_1, V_3, X, Z)$, and $Z \indep (V_1, V_2, X, Y)$. The subsequent part of the algorithm yields an additional 53,894 constraints total, out of which there are 53,856 inequality constraints, in agreement with \citet[Section 5.1]{Pironio_2011}. Proposition \ref{prop:nontrivial} identifies 53,792 of the inequality constraints as nontrivial; the other 64 are nonnegativity constraints. None of the 38 equality constraints is nontrivial. Besides normalization, they are of type (O) and follow from the independence relations above: they state that the distribution of any two of $V_1, V_2, V_3$ does not depend on the setting of the third party, as in $(V_1, V_3) \indep Y \mid X, Z$, $(V_1, V_2) \indep Z \mid X, Y$, and $(V_2, V_3) \indep X \mid Y, Z$. We write $\Pr(v_1 v_2 v_3 \mid xyz) = \Pr(V_1 = v_1, V_2 = v_2, V_3 = v_3 \mid X = x, Y = y, Z = z)$. The district factor of Proposition \ref{prop:ident} is $\Pr(v_1 \mid x)\Pr(v_2 \mid v_1, x, y)\Pr(v_3 \mid v_1, v_2, x, y, z) = \Pr(v_1 v_2 v_3 \mid xyz)$ under these independencies. An example of one of the inequality constraints is
\begin{align*}
& \Pr(001 \mid 010) + \Pr(010 \mid 101) - \Pr(110 \mid 001) - \Pr(010 \mid 100) - \\
& \Pr(000 \mid 111) - \Pr(001 \mid 000) - \Pr(011 \mid 000) \leq 0.
\end{align*}

\subsection{Graphical structures that imply inequality constraints} \label{sec:structures}

The examples in this section and in Section \ref{sec:nontriv} have nontrivial inequality constraints in some districts and none in others. For a district of c-degree 1, two structures in the graph guarantee nontrivial inequality constraints. 

\begin{remark}
\label{prop:ineqstructure}
Let $\mathcal{G}$ satisfy Conditions \ref{A1} and \ref{A2} and have c-degree 1, and let $D$ be a district of $\mathcal{G}$ whose members and external parents each have at least two levels. Call a directed path a \emph{path in $D$} if all of its vertices other than the first are members of $D$. Suppose that $D$ contains at least one of the following:
\begin{enumerate}
\item[(i)] (\emph{instrumental type}) members $A$ and $B$, with $A$ a parent of $B$, and a path in $D$ from some $E \in \bV_2$ to $A$, no vertex of which other than $A$ is a parent of $B$;
\item[(ii)] (\emph{Bell type}) members $A$ and $B$ and external parents $E_A, E_B \in \bV_2$, with paths in $D$ from $E_A$ to $A$ and from $E_B$ to $B$, but none from $E_A$ to $B$ or from $E_B$ to $A$.
\end{enumerate}
Then the H-representation of $D$ contains a nontrivial inequality constraint.
\end{remark}

The remark states that the classical instrumental and CHSH inequalities persist within a district of c-degree 1, at any cardinalities, as nontrivial rows of the H-representation returned by the algorithm. The proof, in Supplementary Materials \ref{appendix:proofs}, deletes edges to reduce $D$ to an instrumental variable or Bell district, which can only shrink the district polytope \citep[Lemma 3]{ansanelli2025observationalpartialordercausal}. The exclusion conditions then ensure that the instrumental or CHSH inequality of the smaller district still holds on the polytope of $D$, so a distribution that violates it, and satisfies the equality constraints of the smaller district, satisfies those of $D$ but lies outside its polytope.

In both types, the inequality constraints come from exclusion restrictions. In type (i), $E$ and the members on the path to $A$ are excluded from $B$, so $E$ is an instrument for the effect of $A$ on $B$. This type appears in the instrumental variable model, in the districts $\{V_2, V_3\}$ and $\{V_4, V_5\}$ of Figure \ref{fig:seqiv}, and in the right graph of Figure \ref{fig:nested-ineq}, with $(E, A, B) = (V_1, V_2, V_4)$. In type (ii), $E_A$ is excluded from $B$ and $E_B$ from $A$. This type appears in the tripartite Bell district, in $X \rightarrow A \leftarrow U \rightarrow B \leftarrow Y$, and in the district $\{V_2, V_4\}$ of $V_1 \rightarrow V_2 \rightarrow V_3 \rightarrow V_4$ with $U \rightarrow V_2$ and $U \rightarrow V_4$. The merged second district of Figure \ref{fig:incomp} contains both types. The districts in our examples without nontrivial inequality constraints contain neither: the front door model, the left graph of Figure \ref{fig:nested-ineq} and the first district of Figure \ref{fig:incomp}. The proof gives one inequality for each type: the CHSH inequality for type (ii), and for type (i),
\begin{equation}
\Pdo(A = a, \boldsymbol{T} = t, B = b \mid \bV_2 = v_2) + \Pdo(A = a, \boldsymbol{T} = t, B = b' \mid \bV_2 = v_2') \leq 1, \label{eqn:ivtype}
\end{equation}
where $b \neq b'$, $\boldsymbol{T}$ are the other members that are parents of $B$, and $v_2$ and $v_2'$ differ only in $E$. These do not exhaust the inequalities; for example, instrumental-type districts acquire new inequalities when the instrument has three or more levels. Two existing criteria, both of which apply to graphs of any c-degree, detect the instrumental type. The first is the e-separation criterion of \citet[Theorem 4.2]{evans2012graphical}: in the instrumental variable model, $Z$ and $Y$ are e-separated after deleting $X$, and in the right graph of Figure \ref{fig:nested-ineq}, $V_1$ and $V_4$ are e-separated after deleting $\{V_2, V_3\}$. The second is the result of \citet{khanna2024classifying} on nonmaximal graphs, because $E$ and $B$ are nonadjacent but cannot be d-separated: conditioning on a member on the path from $E$ to $A$ opens a path through the shared unobserved variable. Neither criterion detects the Bell type. In $X \rightarrow A \leftarrow U \rightarrow B \leftarrow Y$, $A$ and $B$ share the unobserved $U$, which cannot be deleted, and the pairs that can be separated, such as $X$ and $B$, are only marginally independent \citep[Remark 2.6]{evans2023latent}.

Remark \ref{prop:ineqstructure} gives only a sufficient condition, and both structures are known to produce inequality constraints. Whether a graph implies any inequality constraints at all has been studied for hidden variables DAGs in general. \citet{evans2023latent} shows that a discrete hidden variables DAG implies no nontrivial inequality constraints if and only if its marginal model is that of a DAG without hidden variables, and gives the CHSH inequalities of the district $\{V_2, V_4\}$ above as an example. However, it is an open question if existing theory on graph transformations that preserve the marginal model (Proposition \ref{prop:hlp} in the next Section) can be used to obtain the DAG that must exist from the hidden variables DAG. \citet{khanna2024classifying} discuss this conjecture and use the results of \citet{evans2023latent}, together with d-separation, e-separation and incompatible supports, to classify graphs with up to four observed variables, leaving 3 of the 2809 graphs with four observed variables unresolved.

\section{Graphs with c-degree greater than 1}
\label{sec:nonlin}

\subsection{Transformations that may reduce the c-degree} \label{graphmod}

Different graphs can have the same marginal model. For example, the graph $X \rightarrow Y$ has the same marginal model as the graph in which $X$ and $Y$ share an unobserved parent. Such graphs are \emph{observationally} equivalent, although they need not be causally equivalent. Since our method gives complete constraints for graphs with c-degree 1, it is useful to replace a graph by an observationally equivalent graph with a lower c-degree. We use two rules for this. The first, due to \citet[Theorem 26.4; see also Proposition 4 of \citealp{ansanelli2025observationalpartialordercausal}]{henson2014theory}, adds or removes an edge.

\begin{proposition}[HLP Proposition] \label{prop:hlp}
  Let $V_1$ and $V_2$ be observed vertices, and let $\mathcal{G}$ and $\mathcal{G}^+$ be DAGs that meet Conditions \ref{A1} and \ref{A2} and are identical except that $\mathcal{G}^+$ contains the edge $V_1 \to V_2$ and $\mathcal{G}$ does not.
  If:
  \begin{itemize}
      \item $\Pa(V_1) \subseteq \Pa(V_2)$;
      \item There exists a $U_1$ such that $\{V_1, V_2\} \subseteq \Chobs(U_1)$;
  \end{itemize}
  then $\mathcal{G}$ and $\mathcal{G}^+$ have the same constraints. Hence the edge $V_1 \to V_2$ may be added to $\mathcal{G}$ or removed from $\mathcal{G}^+$.
\end{proposition}

The two conditions hold in $\mathcal{G}$ if and only if they hold in $\mathcal{G}^+$, because only the parent set of $V_2$ differs between the two graphs. The second rule is the Strong Facet-Adding rule of \citet[Proposition 8; see also Proposition 15]{ansanelli2025observationalpartialordercausal}, translated into our notation. We state it in the direction that splits unobserved variables, which can reduce the c-degree.

\begin{proposition}[Strong Face Splitting] \label{prop:strong}
  Let $\mathcal{G}$ be a hidden variables DAG satisfying Condition \ref{A1}, with unobserved vertices $U_1, \ldots, U_k$. Let $C_1, \ldots, C_k$ and $C_0$ be sets of observed vertices such that $\Chobs(U_i) = C_i \cup C_0$ for $i = 1, \ldots, k$, and $C_0$ is nonempty and disjoint from every $C_i$; the sets $C_i$ need not be disjoint from each other. Suppose that:
  \begin{enumerate}
      \item $\Paobs(C_i) \cup C_i \subseteq \Paobs(c)$ for each $c \in C_0$ and $i = 1, \dots, k$;
      \item for every unobserved vertex $U_j$ with $j \notin \{1, \ldots, k\}$ such that $\Chobs(U_j)$ intersects $\bigcup_{i=1}^k C_i$ and is not contained in any $C_i$, we have $C_0 \subseteq \Chobs(U_j)$.
  \end{enumerate}
  Define the \emph{split} graph $\mathcal{G}^*$ that has the same nodes and edges as $\mathcal{G}$, except that $U_1, \ldots, U_k$ and their edges are replaced by unobserved vertices $U^*_1, \ldots, U^*_{k+1}$ with $\Chobs(U^*_i) = C_i$ for $i = 1, \ldots, k$ and $\Chobs(U^*_{k+1}) = C_0$. Then $\mathcal{G}$ and $\mathcal{G}^*$ have the same constraints. Hence $U_1, \ldots, U_k$ in $\mathcal{G}$ may be split into $U^*_1, \ldots, U^*_{k+1}$ (face splitting).
\end{proposition}

To apply the proposition, choose $U_1, \ldots, U_k$ with at least one child in common, let $C_0$ be a nonempty subset of $\bigcap_i \Chobs(U_i)$, and let $C_i = \Chobs(U_i) \setminus C_0$. An unobserved vertex of $\mathcal{G}^*$ with a single child can be deleted, so after splitting, the absorption step of Section \ref{sec:settings} restores Condition \ref{A2}. Together with Proposition \ref{prop:hlp}, Proposition \ref{prop:strong} includes the rule for splitting an unobserved vertex in Proposition 5 of \citet{evans2016graphs}.

Figure \ref{fig:ivequiv} shows an example. The left DAG has c-degree 2. Since $\Pa(Z) = \{U_1\} \subseteq \Pa(X)$ and $U_1$ is a parent of both, Proposition \ref{prop:hlp} adds the edge $Z \rightarrow X$, which gives the center DAG. Proposition \ref{prop:strong} then splits $U_1$, with $k = 1$, $C_0 = \{X\}$ and $C_1 = \{Z\}$. The first condition holds because $\Paobs(Z) \cup \{Z\} = \{Z\} \subseteq \Paobs(X)$, and the second because $U_2$ has no child in $C_1$. Both resulting unobserved vertices have a single child and are deleted, which leaves the instrumental variable DAG on the right. It has c-degree 1, so our method gives the complete constraints of all three DAGs.

\begin{figure}[ht]
\centering
    \begin{tikzpicture}
    \begin{scope}[xshift=-2cm]
    \node[draw, circle] at (2, 0) (X) {\(X\)};
    \node[draw, circle] at (4, 0) (Y) {\(Y\)};
    \node[draw, circle] at (0, 0) (Z) {$Z$};
    \node[draw, circle, dashed] at (1,2) (U1) {\(U_1\)};
    \node[draw, circle, dashed] at (3,2) (U2) {\(U_2\)};
    \draw [-{Latex[scale=2.0]}] (X) to (Y);
    \draw [-{Latex[scale=2.0]}] (U2) to (X);
    \draw [-{Latex[scale=2.0]}] (U2) to (Y);
    \draw [-{Latex[scale=2.0]}] (U1) to (X);
    \draw [-{Latex[scale=2.0]}] (U1) to (Z);
    \draw[dashed] (4.75, -1) -- (4.75, 3);
    \end{scope}
    \begin{scope}[xshift=3.5cm]
    \node[draw, circle] at (2, 0) (X) {\(X\)};
    \node[draw, circle] at (4, 0) (Y) {\(Y\)};
    \node[draw, circle] at (0, 0) (Z) {$Z$};
    \node[draw, circle, dashed] at (1,2) (U1) {\(U_1\)};
    \node[draw, circle, dashed] at (3,2) (U2) {\(U_2\)};
    \draw [-{Latex[scale=2.0]}] (X) to (Y);
    \draw [-{Latex[scale=2.0]}] (Z) to (X);
    \draw [-{Latex[scale=2.0]}] (U2) to (X);
    \draw [-{Latex[scale=2.0]}] (U2) to (Y);
    \draw [-{Latex[scale=2.0]}] (U1) to (X);
    \draw [-{Latex[scale=2.0]}] (U1) to (Z);
    \draw[dashed] (4.75, -1) -- (4.75, 3);
    \end{scope}
    \begin{scope}[xshift=9cm]
    \node[draw, circle] at (2, 0) (X) {\(X\)};
    \node[draw, circle] at (4, 0) (Y) {\(Y\)};
    \node[draw, circle] at (0, 0) (Z) {$Z$};
    \node[draw, circle, dashed] at (3,2) (U) {\(U\)};
    \draw [-{Latex[scale=2.0]}] (X) to (Y);
    \draw [-{Latex[scale=2.0]}] (Z) to (X);
    \draw [-{Latex[scale=2.0]}] (U) to (X);
    \draw [-{Latex[scale=2.0]}] (U) to (Y);
    \end{scope}
    \end{tikzpicture}
    \caption{\label{fig:ivequiv} Three observationally equivalent DAGs. The left DAG has c-degree 2. Proposition \ref{prop:hlp} adds the edge $Z \rightarrow X$ (center), and face splitting (Proposition \ref{prop:strong}) then gives the instrumental variable DAG, which has c-degree 1 (right).}
\end{figure}

Not every graph can be reduced in this way. In the district $\{V_2, V_4, V_5\}$ of Figure \ref{fig:incomp}, which has c-degree 2, Proposition \ref{prop:hlp} allows no edge to be added or removed, and the first condition of Proposition \ref{prop:strong} fails whenever $U_1$, $U_3$ or both are split. For example, splitting the children $\{V_2, V_4\}$ of $U_1$ would require $\{V_1, V_2\} \subseteq \Paobs(V_4)$ or $\{V_3, V_4\} \subseteq \Paobs(V_2)$. For such graphs, we propose in the next subsection to merge unobserved variables instead, which gives constraints that hold but need not be complete.

\subsection{Merging hidden variables}

We extend our algorithm to DAGs with c-degree greater than 1 that cannot be reduced to c-degree 1 using the propositions of Section \ref{graphmod}. Prior to applying the algorithm in Section \ref{algorithm}, for each district with c-degree at least 2, we replace all its unobserved variables by a single unobserved parent of all its members.

\begin{remark}
\label{prop:merge}
Let $\mathcal{G}'$ be a DAG satisfying Conditions \ref{A1}, \ref{A2}, and with c-degree greater than 1. Let $\mathcal{G}$ be the causal model obtained from $\mathcal{G}'$ by merging unobserved variables so that the c-degree is 1. 
Then the constraints admitted by $\mathcal{G}$ hold under $\mathcal{G}'$.
\end{remark}

Step 1 of the algorithm should still be applied to $\mathcal{G}'$, because merging can remove conditional independencies.

\begin{figure}[ht]
\begin{center}
\begin{tikzpicture}[>=Stealth, node distance=2cm, on grid, auto]
\begin{scope}
  \node (A) [rv] {$V_1$};
  \node[draw,circle,dashed] (U2) [below right= of A] {$U_2$};
  \node (B) [above right=of U2, rv] {$V_2$};
  \node (C) [right=of B, rv] {$V_3$};
  \node (D) [right=of C, rv] {$V_4$};
  
  \node (F) [above =of D, rv] {$V_5$};

  \node[draw,circle,dashed] (U1) [above =of B] {$U_1$};
  
  \node[draw,circle,dashed] (U3) [above =of C] {$U_3$};
\node (E) [below=of U2, rv] {$V_6$};
  \draw [-{Latex[scale=2.0]}]  (A) -- (B);
  \draw [-{Latex[scale=2.0]}]  (E) -- (C);
  \draw [-{Latex[scale=2.0]}]  (C) -- (D);
  \draw [-{Latex[scale=2.0]}]  (D) -- (F);
  \draw [-{Latex[scale=2.0]}]  (A) -- (E);
  \draw [-{Latex[scale=2.0]}]  (B) -- (C);

  \draw [-{Latex[scale=2.0]}] (U1) -- (B);
  \draw [-{Latex[scale=2.0]}] (U1) -- (D);
  \draw [-{Latex[scale=2.0]}] (U2) -- (A);
  \draw [-{Latex[scale=2.0]}] (U2) -- (E);
  \draw [-{Latex[scale=2.0]}] (U2) -- (C);
  \draw [-{Latex[scale=2.0]}] (U3) -- (F);
  \draw [-{Latex[scale=2.0]}] (U3) -- (D);
   \draw[dashed] (7.75, -5) -- (7.75, 3);
\end{scope}
\begin{scope}[xshift=8.5cm]
  \node (A) [rv] {$V_1$};
  \node[draw,circle,dashed] (U2) [below right= of A] {$U_2$};
  \node (B) [above right=of U2, rv] {$V_2$};
  \node (C) [right=of B, rv] {$V_3$};
  \node (D) [right=of C, rv] {$V_4$};
  
  \node (F) [above =of D, rv] {$V_5$};

  \node[draw,circle,dashed] (U13) [above =of C] {$U_{13}$};
\node (E) [below=of U2, rv] {$V_6$};
  \draw [-{Latex[scale=2.0]}]  (A) -- (B);
  \draw [-{Latex[scale=2.0]}]  (E) -- (C);
  \draw [-{Latex[scale=2.0]}]  (C) -- (D);
  \draw [-{Latex[scale=2.0]}]  (D) -- (F);
  \draw [-{Latex[scale=2.0]}]  (A) -- (E);
  \draw [-{Latex[scale=2.0]}]  (B) -- (C);

  \draw [-{Latex[scale=2.0]}] (U13) -- (B);
  \draw [-{Latex[scale=2.0]}] (U13) -- (D);
  \draw [-{Latex[scale=2.0]}] (U2) -- (A);
  \draw [-{Latex[scale=2.0]}] (U2) -- (E);
  \draw [-{Latex[scale=2.0]}] (U2) -- (C);
  \draw [-{Latex[scale=2.0]}] (U13) -- (F);
\end{scope}
\end{tikzpicture}
\end{center}
\caption{Example model with two districts and c-degree 2 (left side), and the model after merging the unobserved variables $U_1$ and $U_3$ to make the c-degree 1 (right side). \label{fig:incomp}}
\end{figure}
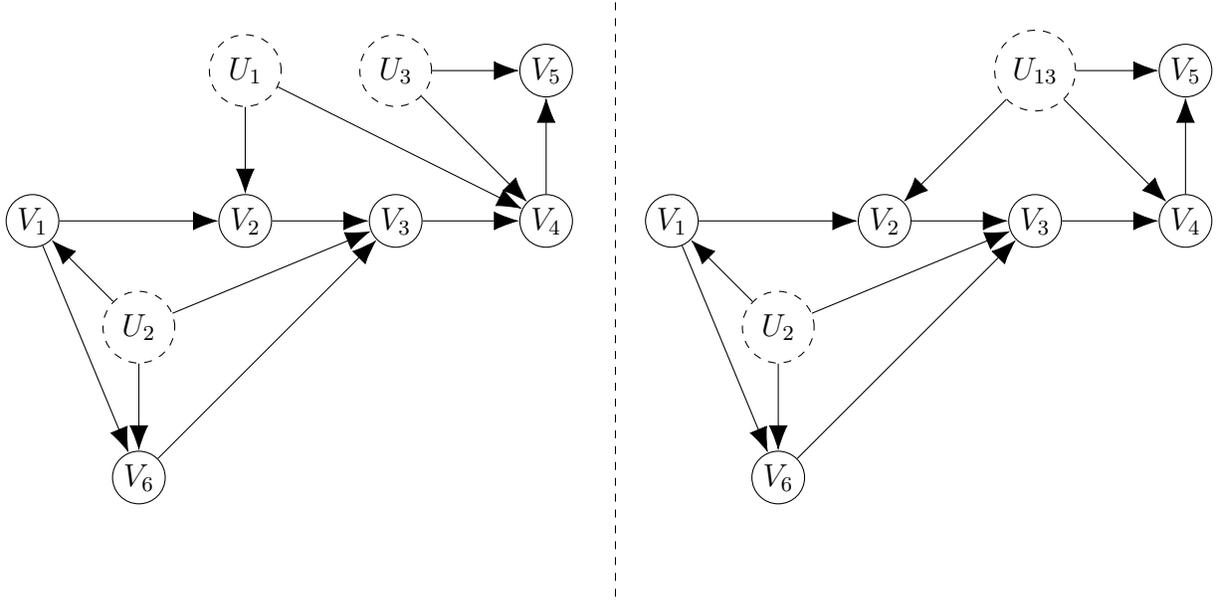

We now consider the two examples where our algorithm does not apply as discussed in Section \ref{sec:complete}.
Figure \ref{fig:incomp}, left side, shows an example DAG that also appears in \citet{duarte2023automated}. This DAG has two districts, $\{V_1, V_3, V_6, U_2\}$ that has c-degree 1 and $\{V_2, V_4, V_5, U_1, U_3\}$ that has c-degree 2. It satisfies Conditions \ref{A1} and \ref{A2}, but since the c-degree is greater than 1, we cannot guarantee that the constraints derived by our algorithm are complete. In this case, we proceed by merging the unobserved variables $U_1$ and $U_3$ into a single unobserved variable $U_{13}$ that has child set equal to the union of the child sets of $U_1$ and $U_3$ (Figure \ref{fig:incomp}, right side).

For the first district in the graph in Figure \ref{fig:incomp} that contains $V_1, V_3, V_6$, Proposition \ref{prop:nontrivial} finds no nontrivial constraints. Besides normalization, the equality constraints are
\[
\Pdo(V_1 = v_1, V_6 = v_6 \mid V_2 = 0) = \Pdo(V_1 = v_1, V_6 = v_6 \mid V_2 = 1), \quad v_1, v_6 \in \{0, 1\},
\]
which are of type (O) with $\bS = \{V_1, V_6\}$. The district factor is $\Pdo(v_1, v_3, v_6 \mid v_2) = \Pr(v_1)\Pr(v_6 \mid v_1)\Pr(v_3 \mid v_1, v_6, v_2)$, so $\Pdo(v_1, v_6 \mid v_2) = \Pr(v_1, v_6)$. The equalities therefore hold identically, and the inequalities reduce to nonnegativity.

After merging $U_1$ and $U_3$ in the second district and applying our algorithm, Proposition \ref{prop:nontrivial} identifies 1096 nontrivial inequality constraints and 6 linearly independent nontrivial equality constraints. The district factor is
\[
\Pdo(v_2, v_4, v_5 \mid v_1, v_3) = \Pr(v_2 \mid v_1)\Pr(v_4 \mid v_1, v_2, v_3)\Pr(v_5 \mid v_1, v_2, v_3, v_4).
\]
Under $\Do(V_1 = v_1, V_3 = v_3)$, the variable $V_2$ depends only on $v_1$ and $U_{13}$, and $V_4$ depends only on $v_3$ and $U_{13}$. Since $V_2$ affects $(V_4, V_5)$ only through $V_3$, which is held fixed, the equality constraints include the condition that $\sum_{v_2}\Pdo(v_2, v_4, v_5 \mid v_1, v_3)$ does not depend on $v_1$. In observed terms this reads
\[
\sum_{v_2} \Pr(v_2 \mid v_1)\Pr(v_4, v_5 \mid v_1, v_2, v_3) \mbox{ does not depend on } v_1,
\]
which is a Verma-type constraint, and which a randomly chosen distribution violates. Together with the trivial equality constraints, this condition implies all 6 nontrivial equality constraints of the district. The set $\{V_4, V_5\}$ is closed in the district but not closed, because $V_2$ is an ancestor of $V_4$ through $V_3$, as in Section \ref{sec:nontriv}. The inequality constraints imply
\begin{equation}
\sum_{v_1, v_3 \in \{0,1\}} \; \sum_{v_2, v_4, v_5 \in \{0,1\}} a(v_1, v_2, v_3, v_4) \, \Pdo(v_2, v_4, v_5 \mid v_1, v_3) \leq 3, \label{eqn:chsh}
\end{equation}
where $a(v_1, v_2, v_3, v_4) = (1 - v_1 v_3)\,\mathbf{1}\{v_2 = v_4\} + v_1 v_3\,\mathbf{1}\{v_2 \neq v_4\}$ and $\mathbf{1}\{\cdot\}$ is the indicator function. Thus $a = 1$ if $V_2$ and $V_4$ are equal when $v_1 v_3 = 0$, or differ when $v_1 = v_3 = 1$, and $a = 0$ otherwise. For each $(v_1, v_3)$, the inner sum in \eqref{eqn:chsh} is the probability of this event under $\Do(V_1 = v_1, V_3 = v_3)$. This is a Bell-type inequality: the four probabilities sum to at most 3. To see that it is nontrivial, consider the following distribution of binary variables:
\begin{itemize}
    \item $\Pr(V_1 = 1) = 1/2$, and $\Pr(V_2 = 1 \mid v_1) = \Pr(V_6 = 1 \mid v_1) = \Pr(V_3 = 1 \mid v_1, v_6, v_2) = \Pr(V_5 = 1 \mid v_1, v_2, v_3, v_4) = 1/2$ for all values of the conditioning variables;
    \item $\Pr(V_4 = v_4 \mid v_1, v_2, v_3) = 1/8 + \frac{3}{4}\, a(v_1, v_2, v_3, v_4)$, that is, $7/8$ if $a(v_1, v_2, v_3, v_4) = 1$ and $1/8$ otherwise.
\end{itemize}
Every probability in the resulting joint distribution of $(V_1, \ldots, V_6)$ is at least $1/256$. The distribution satisfies all the conditional independencies and all the equality constraints of both districts; in particular, $\Pdo(V_4 = 1 \mid v_1, v_3) = 1/2$ for all $v_1, v_3$. However, for each of the four values of $(v_1, v_3)$, the inner sum in \eqref{eqn:chsh} equals $7/8$, so the left side equals $7/2 > 3$. By Remark \ref{prop:merge}, this distribution is therefore also incompatible with the original graph in Figure \ref{fig:incomp}, left. Given any observed distribution, it is straightforward to check the remaining constraints using the matrix representations returned by the software. The example distribution violates 13 of the inequality constraints returned for this district. The shortest of them is
\begin{align*}
\Pdo(V_2 = 0, V_4 = 1, V_5 = 0 \mid V_1 = 1, V_3 = 1) \leq{} & \Pdo(V_2 = 1, V_4 = 0 \mid V_1 = 0, V_3 = 0) \\
& + \Pdo(V_2 = 0, V_4 = 1, V_5 = 0 \mid V_1 = 1, V_3 = 0) \\
& + \Pdo(V_2 = 0, V_4 = 1, V_5 = 0 \mid V_1 = 0, V_3 = 1).
\end{align*}
For the example distribution, the left side equals $7/32$ and the right side equals $1/16 + 1/32 + 1/32 = 1/8$, so the constraint is violated.

Of the 1096 nontrivial inequalities, 16 are instrumental inequalities for $(V_3,V_4,V_5)$ within an event $V_2=v_2$, and 72 are CHSH inequalities for $V_2$ against a binary coarsening of $(V_4,V_5)$. The other 1008 are not of either form, and they are not implied by those two families together with nonnegativity and the equality constraints. Some are closely related to them: the inequality displayed above is a CHSH inequality with one nonnegative term removed, which is valid only because $V_5$ depends on $V_3$ only through $V_4$. If the edge $V_3 \rightarrow V_5$ is added, every nontrivial inequality is a CHSH inequality.

For the triangle scenario, merging the three unobserved variables gives a single common parent of all three observed variables, which has no nontrivial constraints. Thus the constraints are incomplete in this case, because of the nontrivial constraints presented in \citet{Wolfe_2019}.

\section{Implementation} \label{sec:software}

The method is implemented in the R package \texttt{meraconstraints}, available at \if1\blind{\url{https://github.com/sachsmc/meraconstraints}}\fi \if0\blind{\texttt{[blinded url]}}\fi. Graphs are specified using the \texttt{caugi} package, and the V-to-H conversion uses \texttt{cddlib} through the \texttt{rcdd} package. Variables whose names begin with \texttt{U} are treated as latent by default, with the option to specify latent variables in other ways. Code illustrating how to obtain the complete constraints for the instrumental variable model in Figure \ref{fig:ivclassic} is in the Supplementary Material, Section \ref{app:code}, while the other examples in this manuscript are included as R scripts in the ancillary files.

Table \ref{tab:compute} reports the size of the problem and the runtime for a range of districts. For the larger problems the cost is dominated by the V-to-H conversion. Before duplicates are removed, $\boldsymbol{B}$ has one column for each joint value of the response functions, so $\prod_{V \in \bV_1} |V|^{k_V}$ columns, where $k_V$ is the number of configurations of the observed parents of $V$. This number grows exponentially in the number of parent configurations, and so doubly exponentially in the number of parents, and the number of facets can grow faster still. Different joint values of the response functions can give the same column, and these are removed before the V-to-H conversion; this reduces the ternary IV problem from 729 to 333 columns. Identifying the nontrivial constraints with Proposition \ref{prop:nontrivial} takes much less time there (26 s against 1123 s for the tripartite Bell graph).

For a graph with c-degree greater than 1, \texttt{dhvm\_merge\_latents()} carries out the merging step of Section \ref{sec:nonlin} before \texttt{meraconstraints()} is called. R scripts that reproduce the computed results in this paper, including the benchmarks of Table \ref{tab:compute}, are provided as ancillary files with this arXiv submission.

\begin{table}[!htbp]
\centering
\caption{Size of the V-to-H conversion (rows of $\boldsymbol{B}$, and columns of $\boldsymbol{B}$ before and after removing duplicates), number of constraints returned, number of nontrivial inequality constraints and of linearly independent nontrivial equality constraints (Proposition \ref{prop:nontrivial}), and computation time for selected districts. Times are elapsed times in seconds for the V-to-H conversion and for identifying the nontrivial constraints with Proposition \ref{prop:nontrivial}, each for all districts of the graph, on a laptop running Linux (TUXEDO OS 3) with an AMD Ryzen 7 4800H processor and 32 GB of RAM. \label{tab:compute}}
\small
\setlength{\tabcolsep}{4pt}
\begin{tabular}{lrrrrrrrrr}
\hline
 & rows of & \multicolumn{2}{c}{cols of $\boldsymbol{B}$} & \multicolumn{2}{c}{constraints} & \multicolumn{2}{c}{nontrivial} & \multicolumn{2}{c}{time (s)} \\
\cline{3-10}
District & $\boldsymbol{B}$ & all & distinct & ineq. & eq. & ineq. & eq. & V-to-H & Prop.~\ref{prop:nontrivial} \\
\hline
IV, $|Z|=|X|=|Y|=2$ & 8 & 16 & 12 & 12 & 2 & 4 & 0 & $<0.01$ & $<0.01$ \\
IV, $|Z|=3$, $|X|=|Y|=2$ & 12 & 32 & 28 & 48 & 3 & 36 & 0 & $<0.01$ & $<0.01$ \\
IV, $|Z|=4$, $|X|=|Y|=2$ & 16 & 64 & 60 & 160 & 4 & 144 & 0 & 0.13 & 0.04 \\
IV, $|Z|=|X|=|Y|=3$ & 27 & 729 & 333 & 1395 & 3 & 1368 & 0 & 16 & 1.6 \\
Fig.~\ref{fig:nested-ineq} (right), $\{V_2,V_4\}$ & 16 & 64 & 40 & 24 & 6 & 8 & 0 & $<0.01$ & $<0.01$ \\
$X \rightarrow A \leftarrow U \rightarrow B \leftarrow Y$ & 16 & 16 & 16 & 24 & 8 & 8 & 0 & $<0.01$ & $<0.01$ \\
Fig.~\ref{fig:incomp}, merged $\{V_2,V_4,V_5\}$ & 32 & 64 & 48 & 1128 & 12 & 1096 & 6 & 0.22 & 0.22 \\
Tripartite Bell, Fig.~\ref{fig:tripart} & 64 & 64 & 64 & 53,856 & 38 & 53,792 & 0 & 1123 & 26 \\
\hline
\end{tabular}
\end{table}

\clearpage

\section{Discussion and Conclusion}

\citet{bonet2001instrumentality} observed that the complete set of observable constraints in the instrumental variables model can be derived with tools from computational geometry. We have extended this observation to a much broader class of causal models with unobserved variables, showing that a similar set of tools can be used to derive the complete set of equality and inequality constraints for any graph of c-degree 1. Our method uses the response function representation of \citet{balke1994counterfactual}, the conversion from vertices to halfspaces that \citet{bonet2001instrumentality} applied to the instrumental variable model, and the district (c-component) factorization of \citet{tian2002identification, tian2002testable}. \citet{duarte2023automated} use the same response function polynomials to bound causal effects by optimization, whereas we aim to list every restriction that the model places on the observed distribution, without reference to a target quantity.  

We add the following to these known results. We give the observed-data form of the district factor for c-degree 1, and its linear relation to the response function probabilities, so that constraints can be derived one district at a time (Proposition \ref{prop:ident}). We prove that the district constraints together with the conditional independencies are complete (Proposition \ref{prop:complete}), which requires the conditioning sets of Proposition \ref{prop:ident} and the ordered local Markov property. We give an exact criterion for which constraints are nontrivial (Proposition \ref{prop:nontrivial}), and we use it to relate the equality constraints to the nested Markov model. We also give a pre-processing step for graphs with higher c-degree and an open-source implementation.

Within a district, the H-representation returned by the double description method is irredundant: each inequality defines a facet of the district polytope, and the equality constraints form a basis \citep{fukuda1996double}. By Proposition \ref{prop:nontrivial}(ii), a nontrivial inequality is also not implied by the conditional independencies together with all other constraints, including those of other districts. Thus the only constraints that can be removed are the trivial ones, which Proposition \ref{prop:nontrivial} identifies. For large problems, the number of constraints can be reduced further by grouping them into orbits under the symmetries of the model, such as relabelling values of a variable or permuting exchangeable nodes. For example, the 53,856 tripartite Bell inequalities fall into 46 symmetry classes, one of which is nonnegativity \citep{Sliwa2003Symmetries, Pironio_2011}.

Because the constraints are derived one district at a time, two candidate graphs need only be compared on the districts where they differ, either in membership, in the parent sets of members, or, when no ordering of the members is consistent with both graphs, in the ordering that defines $\Pdo$. All other districts and their constraints are shared. The comparison also does not require deriving the constraints of both candidate graphs. To find out whether a constraint of graph $\mathcal{G}_2$ distinguishes it from $\mathcal{G}_1$ when the two have the same district factor, one can evaluate it at every column of the matrix $\boldsymbol{B}$ of $\mathcal{G}_1$. If the factors differ, one can sample distributions from $\mathcal{M}(\mathcal{G}_1)$ and check whether they violate it, which can show a difference but cannot rule one out. Both are illustrated for Figure \ref{fig:nested-ineq} in the reproducibility scripts in the ancillary files. For Figure \ref{fig:nested-ineq}, this can be seen from the graphs. The district $\{V_2, V_4\}$ is of instrumental type in the right graph, so it has inequality constraints (Remark \ref{prop:ineqstructure}). In the left graph, $\Pa(V_2) = \{V_1, U_2\} \subseteq \Pa(V_4)$, so Proposition \ref{prop:hlp} adds $V_2 \rightarrow V_4$, and Proposition \ref{prop:strong} with $C_0 = \{V_4\}$ and $C_1 = \{V_2\}$ then removes $U_2$. This leaves a DAG without hidden variables and the same conditional independence, so the left graph has no inequality constraints.

The method is limited to DAGs with finite and discrete observed variables, though no assumptions are made about the state space of the unobserved variables. It relies on each member of a district having finitely many response functions, but two kinds of continuous variables can be handled. First, suppose a continuous variable has no observed children, for example a continuous outcome, and we coarsen it by grouping its values into finitely many categories. The coarsened variable depends on the same parents as the original, and no other structural equation involves it. Our method, applied to the graph with a discrete variable in its place, therefore gives complete constraints on the distribution of the observed variables with the coarsened variable in place of the continuous one. These constraints are also valid for the original model, since every distribution it generates satisfies them after coarsening, but they need not be complete for it. Applied to a discretized outcome, the inequalities of \citet{kedagni2020generalized} are of this type. Second, if the members of a district are discrete but an external parent $Z$ is continuous, as for a continuous instrument, we can apply the method treating $Z$ as if it took $k$ levels $z_1, \ldots, z_k$. The derivation of each resulting constraint does not depend on which values these labels stand for, so it holds for every choice of $k$ values of $Z$, or equivalently for the supremum of its left-hand side over such choices. The inequalities of \citet{kedagni2020generalized} for a continuous instrument have this form.

The V-to-H conversion can be computationally demanding (Section \ref{sec:software}). The precise complexity of the algorithm depends on the connectivity of the DAG, and the method can be computationally simple in graphs with many variables if they reside in different districts with simple connections between districts. It may be possible to adapt multithreaded implementations of computational geometry algorithms to perform this task \citep{Normaliz}, or to extend the pruning approach of \citet{shridharan2023scalable} to this setting to reduce the size of the conversion in certain cases.

This method has potential applications in causal discovery with hidden variables, as this provides a finer stratification of causal models into equivalence classes than does independence testing or equality constraints alone. How to most effectively incorporate these constraints in causal discovery algorithms is an open question. These constraints also have potential applications in constrained statistical inference and partial identification.

\if1\blind
{
\section*{Acknowledgements}

During the revision our coauthor Professor Robin Evans fell ill and passed away. His spirit lives on in our hearts and minds. 
}
\fi

\bibliographystyle{abbrvnat}
\bibliography{references}

\clearpage
\begin{center}
  {\LARGE\bf Supplementary Materials}
\end{center}
\bigskip

\appendix

\section{Examples of transformations that do not change the marginal model} \label{appendix:trans}

The examples in the following Figures are adapted from Figures 4 and 5 from \citet{evans2016graphs}.

\begin{figure}[ht]
\centering

    \begin{tikzpicture}
    \begin{scope}
    \node[draw, circle] at (2, 0) (X) {\(X\)};
    \node[draw, circle] at (4, 0) (Y) {\(Y\)};
    \node[draw, circle] at (0, 0) (Z) {$Z$};
    \node[draw, circle, dashed] at (4,2) (U) {\(U_1\)};
    \node[draw, circle, dashed] at (2,2) (U2) {\(U_2\)};
    
    \draw [-{Latex[scale=2.0]}] (Z) to (U2);
    \draw [-{Latex[scale=2.0]}] (U2) to (U);
    \draw [-{Latex[scale=2.0]}] (Z) to (X);
    \draw [-{Latex[scale=2.0]}] (U) to (X);
    \draw [-{Latex[scale=2.0]}] (U) to (Y);
    \draw[dashed] (4.75, -1) -- (4.75, 3);

    \end{scope}
\begin{scope}[xshift=5.5cm]
  \node[draw, circle] at (2, 0) (X) {\(X\)};
    \node[draw, circle] at (4, 0) (Y) {\(Y\)};
    \node[draw, circle] at (0, 0) (Z) {$Z$};
    \node[draw, circle, dashed] at (3,2) (U) {\(U\)};
    
    \draw [-{Latex[scale=2.0]}, bend right] (Z) to (Y);
    \draw [-{Latex[scale=2.0]}] (Z) to (X);
    \draw [-{Latex[scale=2.0]}] (U) to (X);
    \draw [-{Latex[scale=2.0]}] (U) to (Y);
    \end{scope}
    \end{tikzpicture}
    \caption{An example of the exogenization process. The two graphs induce the same marginal model over $\{Z, X, Y\}$. The graph on the right side satisfies Condition \ref{A1}, but the graph on the left does not. }
    \end{figure}
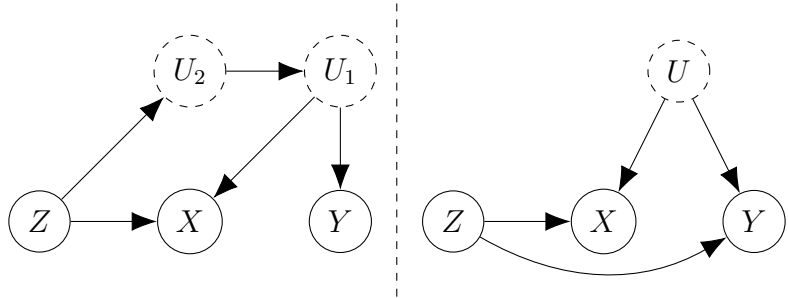

    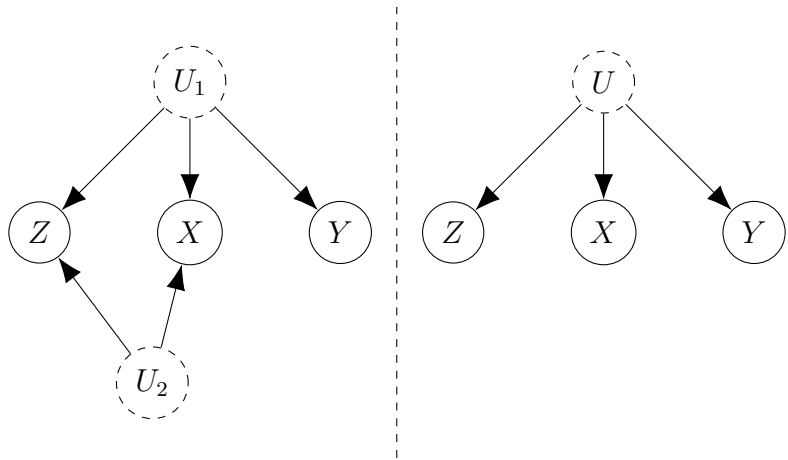
\begin{figure}[ht]
\centering

    \begin{tikzpicture}
    \begin{scope}
    \node[draw, circle] at (2, 0) (X) {\(X\)};
    \node[draw, circle] at (4, 0) (Y) {\(Y\)};
    \node[draw, circle] at (0, 0) (Z) {$Z$};
    \node[draw, circle, dashed] at (2,2) (U) {\(U_1\)};
    \node[draw, circle, dashed] at (1.5,-2) (U2) {\(U_2\)};
    
    \draw [-{Latex[scale=2.0]}] (U2) to (Z);
    \draw [-{Latex[scale=2.0]}] (U2) to (X);
    \draw [-{Latex[scale=2.0]}] (U) to (X);
    \draw [-{Latex[scale=2.0]}] (U) to (Y);
    \draw [-{Latex[scale=2.0]}] (U) to (Z);
    
    \draw[dashed] (4.75, -3) -- (4.75, 3);

    \end{scope}
\begin{scope}[xshift=5.5cm]
 \node[draw, circle] at (2, 0) (X) {\(X\)};
    \node[draw, circle] at (4, 0) (Y) {\(Y\)};
    \node[draw, circle] at (0, 0) (Z) {$Z$};
    \node[draw, circle, dashed] at (2,2) (U) {\(U\)};
    
    \draw [-{Latex[scale=2.0]}] (U) to (X);
    \draw [-{Latex[scale=2.0]}] (U) to (Y);
    \draw [-{Latex[scale=2.0]}] (U) to (Z);
    \end{scope}
    \end{tikzpicture}
    \caption{An example of the ``nested unobserved variable absorbing'' process. The two graphs induce the same marginal model over $\{Z, X, Y\}$. The graph on the right side satisfies Condition \ref{A2}, but the graph on the left does not. }
    \end{figure}  

\clearpage

\section{Nested Markov model} \label{sec:nmm}

We summarize the nested Markov model, following \citet{richardson2023nested}. A \emph{kernel} $q_{\bS}(v_{\bS} \mid v_{\bV \setminus \bS})$ is a distribution over $\bS$ for each value of the remaining variables, which are treated as fixed; in the associated graph, the edges into the fixed variables are removed. A variable $V \in \bS$ is \emph{fixable} if no other member of its district in the current graph is a descendant of $V$. Fixing $V$ divides the kernel by the conditional distribution of $V$ given the other members of its district and the parents of the district, and moves $V$ to the fixed variables. A set $\bS$ is \emph{reachable} if it can be obtained from $\bV$ by fixing one variable at a time. A distribution is in the nested Markov model if every reachable kernel obeys the global Markov property of its graph. In the discrete case, these are the only equality constraints of the marginal model \citep{evans2018margins}.

In a district of c-degree 1, the district factor is the kernel reached by fixing $\bV \setminus \bV_1$ in reverse topological order. Any two unfixed members share a district, so a member can be fixed only when it has no children among the remaining members. Hence, whatever the order of fixing, the reachable subsets of $\bV_1$ are exactly the sets $\bS$ closed in the district, with kernel the margin $p_{\bS}$ of Section \ref{sec:nontriv}. Each such kernel has a single complete district, so its global Markov property states only that it does not depend on $\bV_2 \setminus \boldsymbol{E}(\bS)$. The example $\sum_B \Pr(D \mid A, B, C)\Pr(B \mid A) = g(D, C)$ of Section \ref{sec:nested} is such a margin, with $B$ fixed.

\section{Details of instrumental variable example} \label{appendix:iv}

The algorithm applied to the second district which contains $X$ and $Y$ yields the matrix $\boldsymbol{B}=$
\[
\begin{pmatrix}
 1 & 1 & 1 & 1 & 0 & 0 & 0 & 0 & 0 & 0 & 0 & 0 & 0 & 0 & 0 & 0 \\ 
 0 & 0 & 0 & 0 & 1 & 1 & 1 & 1 & 0 & 0 & 0 & 0 & 0 & 0 & 0 & 0 \\ 
 0 & 0 & 0 & 0 & 0 & 0 & 0 & 0 & 1 & 1 & 1 & 1 & 0 & 0 & 0 & 0 \\ 
 0 & 0 & 0 & 0 & 0 & 0 & 0 & 0 & 0 & 0 & 0 & 0 & 1 & 1 & 1 & 1 \\ 
 1 & 0 & 1 & 0 & 1 & 0 & 0 & 0 & 0 & 0 & 0 & 0 & 1 & 0 & 0 & 0 \\ 
 0 & 1 & 0 & 0 & 0 & 1 & 0 & 1 & 0 & 1 & 0 & 0 & 0 & 0 & 0 & 0 \\ 
 0 & 0 & 0 & 0 & 0 & 0 & 1 & 0 & 1 & 0 & 1 & 0 & 0 & 0 & 1 & 0 \\ 
 0 & 0 & 0 & 1 & 0 & 0 & 0 & 0 & 0 & 0 & 0 & 1 & 0 & 1 & 0 & 1 
\end{pmatrix}
\]
where the rows correspond to the constraints on the vector of probabilities $[\Pr(Y = y \mid X = x, Z = z) \Pr(X = x \mid Z = z)]$ for 
\[
(x, y, z) = 
\begin{pmatrix}
0 & 0 & 0 \\ 
 1 & 0 & 0 \\ 
 0 & 1 & 0 \\ 
 1 & 1 & 0 \\ 
 0 & 0 & 1 \\ 
 1 & 0 & 1 \\ 
 0 & 1 & 1 \\ 
 1 & 1 & 1 
\end{pmatrix}.
\]
After adding the row of 1s to the $\boldsymbol{B}$ matrix and doing the V-to-H conversion, we obtain the H-representation defined by $b_i = (0, 0, 0, 0, 1, 1, 0, 1, 0, 1, 0, 0)^\top$,
\[
H_i = \begin{pmatrix}
0 & 0 & 0 & 0 & -1 & 0 & 0 & 0 \\
 -1 & -1 & -1 & 0 & 0 & 1 & 0 & 0 \\
 0 & 1 & 0 & 0 & -1 & -1 & -1 & 0 \\
 -1 & 0 & 0 & 0 & 0 & 0 & 0 & 0 \\
 0 & 0 & 1 & 0 & 1 & 0 & 0 & 0 \\
 1 & 0 & 0 & 0 & 0 & 0 & 1 & 0 \\
 0 & 0 & 0 & 0 & 0 & -1 & 0 & 0 \\
 0 & 0 & 0 & 0 & 1 & 1 & 1 & 0 \\
 0 & -1 & 0 & 0 & 0 & 0 & 0 & 0 \\
 1 & 1 & 1 & 0 & 0 & 0 & 0 & 0 \\
 0 & 0 & 0 & 0 & 0 & 0 & -1 & 0 \\
 0 & 0 & -1 & 0 & 0 & 0 & 0 & 0 \\
\end{pmatrix},
\]
$b_e = (-1, -1)^\top$, and 
\[
H_e = \begin{pmatrix}
 -1 & -1 & -1 & -1 & 0 & 0 & 0 & 0 \\
 0 & 0 & 0 & 0 & -1 & -1 & -1 & -1 \\
\end{pmatrix}.
\]

Rows 1, 4, 7, 8, 9, 10, 11 and 12 of $H_i$ are nonnegativity constraints, rows 8 and 10 after adding a normalization constraint, and the equality constraints are the normalization constraints. By Proposition \ref{prop:nontrivial}, the other 4 rows of $H_i$ are nontrivial.

\section{Handling events of probability zero} \label{app:positivity}

Proposition \ref{prop:complete} extends to distributions with zero probabilities as follows. Fix a topological ordering of $\mathcal{G}$, order every district by it, and read each product in Proposition \ref{prop:ident} from $i = 1$ to $k$. The first factor that is not positive decides the entry of $\boldsymbol{p}$. If it is zero, the entry is zero regardless of the later factors. If it is undefined, because $\Pr(\boldsymbol{C}_i = c_i) = 0$, the observed distribution does not determine the entry, and we treat the entry as a free variable. Then $\Pr \in \mathcal{M}(\mathcal{G})$ if and only if $\Pr$ satisfies the conditional independencies and, for each district, the free entries can be chosen so that $H_i \boldsymbol{p} \leq b_i$ and $H_e \boldsymbol{p} = b_e$. Equivalently, there is a probability vector $\boldsymbol{r}$ with $\boldsymbol{p} = \boldsymbol{B}\boldsymbol{r}$ on the remaining entries, which is a linear program. A strictly positive distribution has no free entries. Which entries are free can depend on the topological ordering, but the equivalence holds for every topological ordering.

Since $\boldsymbol{C}_i \subseteq \boldsymbol{C}_{i+1}$ and $V_{1i} \in \boldsymbol{C}_{i+1}$, a factor that follows a zero or undefined factor is undefined, so the rule above is well defined. By the equivalence of the V- and H-representations, the free entries can be chosen so that the constraints of a district hold if and only if there is a probability vector $\boldsymbol{r}$ with $\boldsymbol{p} = \boldsymbol{B}\boldsymbol{r}$ on the other entries. Fix the topological ordering.

The matrices $\boldsymbol{B}$, $H_i$ and $H_e$ do not depend on the ordering, because the rows of $\boldsymbol{B}$ are indexed by the values $(v_1, v_2)$. Only the map from $\Pr$ to the entries of $\boldsymbol{p}$ does. Fix any topological ordering of $\mathcal{G}$, and compute the entries of every district with the ordering that it induces on the district. The argument below uses only that this ordering is topological and that the same ordering is used for every district. The equivalence therefore holds for each topological ordering, and whether its condition holds does not depend on the ordering chosen, although the set of free entries can.

First suppose that $\Pr \in \mathcal{M}(\mathcal{G})$. Then $\Pr$ satisfies the conditional independencies \citep{evans2016graphs}, and, as in the proof of Proposition \ref{prop:complete}, $\Pr(v) = \prod_{D'} p_{D'}(v_{D'} \mid v_{\bV_2(D')})$, where each $\boldsymbol{p}_{D'} = \boldsymbol{B}_{D'} \boldsymbol{r}_{D'}$ is the interventional distribution of $D'$ under the model, which does not depend on the ordering. Each $\boldsymbol{p}_{D'}$ lies in its district polytope and hence satisfies (N) and (O). The derivation of \eqref{eqn:ancmargin} uses only (N) and (O), so that equation holds here. Let $\boldsymbol{m}$ denote the margins of $\boldsymbol{p}_D$ for a district $D$. Apply \eqref{eqn:ancmargin} to the variables up to and including $V_{1i}$ in the fixed ordering, and to those before $V_{1i}$. Both sets contain the observed ancestors of their elements, because the ordering is topological. The factors of the other districts are the same in both, and we call their product $Q$, so that the two probabilities are $m_{\bV_1^{\leq i}} Q$ and $m_{\bV_1^{<i}} Q$. Suppose that $\Pr(\boldsymbol{C}_i = c_i) > 0$. Then $m_{\bV_1^{<i}} > 0$ at $c_i$. Both margins depend only on $v_{1i}$ and $c_i$, so summing over the variables outside $\boldsymbol{C}_i \cup \{V_{1i}\}$ gives $\Pr(v_{1i} \mid c_i) = m_{\bV_1^{\leq i}} / m_{\bV_1^{<i}}$. These ratios telescope. If the first factor that is not positive is the $l$th and it is zero, then $m_{\bV_1^{\leq l}} = 0$, so the entry of $\boldsymbol{p}_D$ is zero. If all factors are positive, their product is the entry of $\boldsymbol{p}_D$. Thus $\boldsymbol{p}_D$ agrees with every entry that is not free.

Conversely, suppose that $\Pr$ satisfies the conditional independencies and that, for each district, $\boldsymbol{p}_{D'} = \boldsymbol{B}_{D'} \boldsymbol{r}_{D'}$ agrees with every entry that is not free. As in the proof of Proposition \ref{prop:complete}, the distribution $\prod_{D'} p_{D'}(v_{D'} \mid v_{\bV_2(D')})$ is in $\mathcal{M}(\mathcal{G})$, so it suffices to show that it equals $\Pr(v)$ for every $v$. If $\Pr(v) > 0$, every conditioning event in the chain rule has positive probability. The ordered local Markov property replaces each conditioning set by $\boldsymbol{C}_V$, as in that proof, and every factor is positive, so $\Pr(v)$ is the product of the entries. If $\Pr(v) = 0$, let $V$ be the first variable in the ordering such that the probability of the values of $v$ up to and including $V$ is zero. The earlier values have positive probability, so the ordered local Markov property gives $\Pr(v_V \mid c_V) = 0$. The earlier members of the district of $V$ come before $V$ in the same ordering, so their conditioning events have positive probability and their factors are positive. Thus the entry of that district is zero, and the product is zero. This is the step that needs one ordering for the chain rule and for the entries of every district. 

For example, in the instrumental variable model with $\Pr(Z = 1) = 0$, every entry with $z = 1$ is free, and $\Pr(x, y \mid Z = 0)$ is unrestricted. Free entries do not remove all restrictions, however. In the graph $A \rightarrow B \rightarrow C \rightarrow D$ with $U \rightarrow B$ and $U \rightarrow D$, if $C = B$ with probability one and $\Pr(b \mid a) > 0$ for all $a, b$, the entries with $c \neq b$ are free, and the other entries must satisfy the instrumental inequalities for $(A, B, D)$.

\section{Code for the IV example} \label{app:code}

\begin{verbatim}
library(meraconstraints)
ivdag  <- caugi(Z %-->% X, X %-->% Y, U %-->% X + Y) |> dhvm()
ivmera <- meraconstraints(ivdag)
ivmera                        # summary of the constraints
H <- ivmera$polytope$Hrep[[1]]
P <- ivmera$functional_relations$probform[[1]]
## all constraints, written in observed probabilities
hrep_to_text(H, P)
## only the nontrivial ones
hrep_to_text(H, P)[ivmera$nontrivial[[1]]]
## test a given vector of district factors
hrep_test(H, c(0.101, 0.338, 0.013, 0.548, 0.317, 0.51, 0.004, 0.169),
          which = TRUE)
\end{verbatim}
The last call reports, for each row of $H_i$ and $H_e$, whether the vector satisfies it. This vector satisfies all of them except the 2nd row of $H_i$, the first of the instrumental inequalities in Section \ref{sec:nontriv}.

\section{Proofs} \label{appendix:proofs}
\subsection*{Proof of Proposition \ref{prop:ident}}

Fix a topological ordering of $\bV$ consistent with the ordering $\bV_1 = (V_{11},\ldots,V_{1k})$ of the district $D_1$; such an ordering exists because the ordering of $\bV_1$ is consistent with a topological ordering of $\mathcal{G}$. Let $\operatorname{pre}(V_{1i})$ denote the set of observed variables preceding $V_{1i}$ in it, and let $\boldsymbol{C}_i = \bV_1^{<i} \cup \{\bV_2 \cap \Paobs(\bV_1^{\leq i})\}$ be as in the statement. We first show that the product defining $\Pdo(\bV_1=v_1\mid \bV_2=v_2)$ is the district (c-component) factor associated with $D_1$ in the factorization of \citet{tian2002identification}. By their Lemma 1, the c-component factor is
\[
Q[D_1](v) = \prod_{i=1}^k \Pr\!\left(V_{1i} = v_{1i} \mid \operatorname{pre}(V_{1i}) = v^{(1i)}\right),
\]
where $v^{(1i)}$ is the subvector of $v$ for the variables in $\operatorname{pre}(V_{1i})$. Let $T_i$ be the district containing $V_{1i}$ in the subgraph of $\mathcal{G}$ induced by $\operatorname{pre}(V_{1i}) \cup \{V_{1i}\}$ and the unobserved variables. Because $D_1$ has c-degree at most 1, its members, if there is more than one, all share the single unobserved parent, so $T_i = \bV_1^{\leq i}$. Hence $\{T_i \cup \Paobs(T_i)\} \setminus \{V_{1i}\} = \boldsymbol{C}_i$, and $V_{1i}$ is independent of $\operatorname{pre}(V_{1i}) \setminus \boldsymbol{C}_i$ given $\boldsymbol{C}_i$ \citep{tian2002testable}. This is the ordered local Markov property \citep{richardson2003admgs}. Since $\boldsymbol{C}_i \subseteq \operatorname{pre}(V_{1i})$, this conditional independence gives
\[
\Pr(v_{1i} \mid v^{(1i)}) = \Pr(v_{1i} \mid c_i),
\]
which proves the claim. The same argument shows that conditioning on any set $\boldsymbol{S}$ with $\boldsymbol{C}_i \subseteq \boldsymbol{S} \subseteq \operatorname{pre}(V_{1i})$ gives the same factor. In general, however, $\boldsymbol{C}_i$ cannot be replaced by $\bV_1^{<i} \cup \{\bV_2 \cap \Anobs(V_{1i})\}$. External parents of earlier members of the district become informative about the shared unobserved variable once those members are conditioned on.

Since $D_1$ has c-degree at most 1, there is at most one unobserved variable whose observed children intersect $D_1$.  By the response function variable transformation described in Section \ref{sec:respvar}, the value of each observed variable $V_{1i}$ in the district may therefore be represented as determined by the value of a finite response function variable $R_{V_{1i}}$ and a deterministic response function
\[
    f^{r_i}_{V_{1i}}: \operatorname{domain}(\Paobs(V_{1i})) \rightarrow \operatorname{domain}(V_{1i}),
\]
so that, for $r_i$ a value of $R_{V_{1i}}$, $V_{1i} = f^{r_i}_{V_{1i}}(\Paobs(V_{1i})).$ Let
\[
    \bR_{\bV_1}
    =
    (R_{V_{11}},\ldots,R_{V_{1k}})
\]
and write $\mathcal{R}_{\bV_1}$ for its finite state space.  The
joint distribution of $\bR_{\bV_1}$ is unrestricted, except for the
usual probability simplex constraints, because the variables in the
same district share an unobserved common cause.

Fix values $v_2$ of $\bV_2$ and $r=(r_1,\ldots,r_k)$ of $\bR_{\bV_1}$.  Since $\mathcal{G}$ is acyclic and the ordering of $\bV_1$ is topological, the structural equations for the variables in $D_1$ may be evaluated recursively. This gives the value $F_{\bV_1}(v_2,r)$ that the variables in the district would take if the external observed parents were fixed to $v_2$ and the response function variables in the district were fixed to $r$.

For each pair $(v_1,v_2)$, define the compatibility set
\[
    \mathcal{B}(v_1,v_2)
    =
    \left\{
        r \in \mathcal{R}_{\bV_1}
        :
        F_{\bV_1}(v_2,r)=v_1
    \right\}.
\]
By construction of the response function representation,
\[
    \Pr(\bV_1=v_1 \mid \Do(\bV_2=v_2))
    =
    \sum_{r\in \mathcal{B}(v_1,v_2)}
    \Pr(\bR_{\bV_1}=r).
\]
The right-hand side is a linear function of the response function
probabilities
\[
    \left\{
        \Pr(\bR_{\bV_1}=r)
        :
        r\in\mathcal{R}_{\bV_1}
    \right\}.
\]

It remains only to connect the observable product defining $\Pdo$ to this interventional quantity. By the first part of the proof, $\Pdo(\bV_1=v_1\mid \bV_2=v_2) = Q[D_1](v)$. By \citet{tian2002identification}, $Q[D_1](v) = \Pr(\bV_1 = v_1 \mid \Do(\bV \setminus \bV_1 = v \setminus v_1))$, and this equals $\Pr(\bV_1 = v_1 \mid \Do(\bV_2 = v_2))$ because the structural equations for $\bV_1$ involve the intervened variables only through $\bV_2$.  Combining these equalities gives
\[
    \Pr(\bV_1=v_1\mid \Do(\bV_2=v_2)) = \Pdo(\bV_1=v_1\mid \bV_2=v_2) = \sum_{r\in \mathcal{B}(v_1,v_2)}
    \Pr(\bR_{\bV_1}=r).
\]

Finally, collecting these equations over all possible values of
$(v_1,v_2)$ gives a matrix representation
\[
    \boldsymbol{p} = \boldsymbol{B} \boldsymbol{r},
\]
where $\boldsymbol{p}$ is the vector with entries
$\Pdo(\bV_1=v_1\mid \bV_2=v_2)$, $\boldsymbol{r}$ is the vector with entries
$\Pr(\bR_{\bV_1}=r)$, and the entries of $\boldsymbol{B}$ are indicators of
membership in the compatibility sets $\mathcal{B}(v_1,v_2)$.  Thus
$\boldsymbol{B}$ is a matrix of zeros and ones, and the functional constraints are linear. This concludes the proof.

\subsection*{Proof of Proposition \ref{causalprop}}
For a district of c-degree 1, the equality of $\Pr(\bV_1 = v_1 \mid \Do(\bV_2 = v_2))$ and $\Pdo(\bV_1 = v_1 \mid \bV_2 = v_2)$ is shown in the proof of Proposition \ref{prop:ident}. For a district of higher c-degree, the set $T_i$ in that proof satisfies $T_i \subseteq \bV_1^{\leq i}$, so $\boldsymbol{C}_i$ contains $\{T_i \cup \Paobs(T_i)\} \setminus \{V_{1i}\}$ and is contained in $\operatorname{pre}(V_{1i})$. By weak union, the conditional independence in that proof still gives $\Pr(v_{1i} \mid v^{(1i)}) = \Pr(v_{1i} \mid c_i)$, and the rest of that argument is unchanged. Identification then follows from Lemma 1, part (i) of \citet{tian2002identification} and the definition of \emph{Causal Effect Identification} in \citet{Pearl2000causality}, Definition 3.2.4, page 77.

\subsection*{Proof of Proposition \ref{prop:complete}}

Let \(D_1,\ldots,D_M\) denote the districts of \(\mathcal{G}\).  For a district \(D_j\), write $\bV^j_1$ for the observed members of $D_j$ and $\bV^j_2 = \Paobs(\bV^j_1)\setminus \bV^j_1$, and let $\boldsymbol{p}_j$ be the vector of all district factors $\Pdo(\bV^j_1=v^j_1\mid \bV^j_2=v^j_2)$, in the ordering used by the algorithm. By Proposition \ref{prop:ident}, $\boldsymbol{p}_j = \boldsymbol{B}_j \boldsymbol{r}_j$, where $\boldsymbol{r}_j$ is the vector of probabilities of all joint states of the response function variables $\bR_{\bV^j_1}$ and $\boldsymbol{B}_j$ is the zero-one compatibility matrix constructed by the algorithm. The joint distribution of the response function variables within a district is unrestricted, apart from the usual probability simplex constraints, while response function variables in distinct districts are independent. Therefore the set of possible district factors for $D_j$ is the convex polytope $\mathcal{P}_j = \{\boldsymbol{B}_j\rho_j : \rho_j\in\Delta_j\}$, where $\Delta_j$ is the probability simplex over $\bR_{\bV^j_1}$. Equivalently, it is the convex hull of the columns of $\boldsymbol{B}_j$, i.e., the vertex- (V-)representation of the district-level model. The vertex-to-halfspace (V-to-H) conversion used by the algorithm returns an equivalent H-representation, so that $\mathcal{P}_j = \{\boldsymbol{p}_j : H^j_{i}\boldsymbol{p}_j \le b^j_{i} \mbox{ and } H^j_{e}\boldsymbol{p}_j = b^j_{e}\}$ \citep{doubledescriptionmethod}.

We now prove the two inclusions, $\mathcal{M}(\mathcal{G}) \subseteq \mathcal{C}(\mathcal{G})$ and $\mathcal{C}(\mathcal{G}) \subseteq \mathcal{M}(\mathcal{G})$.

First suppose that $\Pr\in\mathcal{M}(\mathcal{G})$. Then $\Pr$ is generated by a hidden variables model compatible with $\mathcal{G}$.  Hence the ordinary conditional independencies implied by $\mathcal{G}$ hold \citep{evans2016graphs}.  Further, for each district $D_j$, the response function probabilities $r_j$ induced by the model satisfy
\[
    \boldsymbol{p}_j=\boldsymbol{B}_j r_j.
\]
Hence $\boldsymbol{p}_j$ is in the V-representation used by the algorithm, and therefore satisfies the equivalent H-representation constraints, i.e., $\Pr\in \mathcal{C}(\mathcal{G})$.

Conversely, suppose that $\Pr\in \mathcal{C}(\mathcal{G})$, i.e., it satisfies the conditional independencies and the H-representation constraints returned by the algorithm.  Then, for each district $D_j$, equivalence of the H- and V-representations implies that
there exists a probability vector $r_j$ such that
\[
    \boldsymbol{p}_j=\boldsymbol{B}_j r_j.
\]
Use these $r_j$s as the joint distributions of the response function variables within the corresponding districts, and take response function variables in distinct districts to be mutually independent.  This construction is compatible with \(\mathcal{G}\), since response function variables may be dependent within a district and are independent across districts.

Let $\widetilde{P}$ denote the observed distribution induced by this constructed response function model. For every observed configuration $v$, the distribution induced by the constructed response function model satisfies, with $\mathcal{B}(v^j_1, v^j_2)$ and $\bR_{\bV^j_1}$ defined as in the proof of Proposition \ref{prop:ident},
\begin{align*}
    \widetilde{P}(\bV=v) &= 
    \prod_{j=1}^M
    \left[
        \sum_{r\in\mathcal{B}(v^j_1,v^j_2)}
        \Pr(\bR_{\bV^j_1} = r)
    \right]  \\
    & = 
    \prod_{j=1}^M
    \Pdo(\bV_1^j = v^j_1 \mid \bV_2^j = v^j_2),
\end{align*}
where $\Pdo$ on the right is computed from $\Pr$. The first equality holds because, in the constructed model, the event $\bV = v$ occurs if and only if, for every $j$, $\bR_{\bV_1^j} \in \mathcal{B}(v_1^j, v_2^j)$. This is a consequence of the recursive evaluation in the proof of Proposition \ref{prop:ident}: the value of $\bV_1^j$ is determined by $\bR_{\bV_1^j}$ and $v_2^j$. The probability then factorizes because response function variables in distinct districts are independent. The second equality holds because $\boldsymbol{p}_j = \boldsymbol{B}_j r_j$.

It remains to show that $\Pr(\bV = v) = \prod_{j} \Pdo(\bV_1^j = v^j_1 \mid \bV_2^j = v^j_2)$ for every $v$. Fix a topological ordering of $\bV$. By the chain rule, $\Pr(\bV = v) = \prod_{V \in \bV} \Pr(V = v_V \mid \operatorname{pre}(V) = v^{\operatorname{pre}(V)})$. Grouping the factors by district gives $\prod_j Q_j(v)$, where $Q_j(v) = \prod_{V \in D_j} \Pr(V = v_V \mid \operatorname{pre}(V) = v^{\operatorname{pre}(V)})$. Since $\Pr \in \mathcal{C}(\mathcal{G})$, it satisfies the conditional independencies implied by $\mathcal{G}$. These include the ordered local Markov property \citep{richardson2003admgs}: each $V$ is independent of $\operatorname{pre}(V) \setminus \boldsymbol{C}_V$ given $\boldsymbol{C}_V$, where $\boldsymbol{C}_V$ is the conditioning set of Proposition \ref{prop:ident} for $V$, computed with the ordering of its district induced by the fixed topological ordering. If the algorithm used a different ordering of $D_j$, the factor is unchanged, because under the conditional independencies it does not depend on the ordering (Step 1 of the proof of Proposition \ref{prop:nontrivial}, which does not use the present proposition). As shown in the proof of Proposition \ref{prop:ident}, this allows the conditioning set $\operatorname{pre}(V)$ to be replaced by $\boldsymbol{C}_V$, so that $Q_j(v) = \Pdo(\bV_1^j = v^j_1 \mid \bV_2^j = v^j_2)$. Hence
\[
    \widetilde{P}(\bV=v)=\Pr(\bV=v)
\]
for all $v$, and therefore $\Pr\in\mathcal{M}(\mathcal{G})$. This step requires the conditional independencies: without them, the observed distribution need not equal the product of its district factors.

\subsection*{Proof of Proposition \ref{prop:nontrivial}}

For an ordering $\sigma = (V_{11}, \ldots, V_{1k})$ of $\bV_1$ that is consistent with a topological ordering of $\mathcal{G}$, and a strictly positive distribution $\Pr$ of the observed variables, let $\boldsymbol{p}^\sigma(\Pr)$ be the vector defined in Proposition \ref{prop:ident} using $\sigma$. Thus $\boldsymbol{p} = \boldsymbol{p}^{\sigma_0}(\Pr)$, where $\sigma_0$ is the ordering used by the algorithm. Let $\mathcal{P}_D$ be the district polytope, the convex hull of the columns of $\boldsymbol{B}$, and $\mathcal{A} = \{\boldsymbol{p} : H_e \boldsymbol{p} = b_e\}$ its affine hull. Let $L$ be the set of vectors $\boldsymbol{p}$ that satisfy (N) and (O), and $\boldsymbol{\check{P}}_O = \boldsymbol{\check{P}} \cap L$. In each column of $\boldsymbol{B}$, every member of $D$ is a function of its observed parents. For each member, one of these functions is constant, so every entry $p_k$ of $\boldsymbol{p}$ equals one at some column. In particular, the average $\bar\beta$ of the columns of $\boldsymbol{B}$ is strictly positive. The proof has four steps.

\emph{Step 1: (N) and (O) are trivial, and they hold on $\mathcal{A}$.} 

For every $\sigma$ and every strictly positive $\Pr$, summing $\boldsymbol{p}^\sigma(\Pr)$ over $v_{1k}, v_{1,k-1}, \ldots, v_{1,i+1}$ in turn gives $p^\sigma_{\bV_1^{\leq i}}(v_{\bV_1^{\leq i}} \mid v_2) = \prod_{l \leq i} \Pr(V_{1l} = v_{1l} \mid \boldsymbol{C}_l = c_l)$, which depends on $v_2$ only through its value on $\boldsymbol{E}(\bV_1^{\leq i})$. Thus $\boldsymbol{p}^\sigma(\Pr)$ satisfies (N), and (O) for the closed sets $\bV_1^{\leq i}$ of $\sigma$.

If $\Pr$ also satisfies the conditional independencies, then $\boldsymbol{p}^\sigma(\Pr)$ is the same for every $\sigma$. Any two orderings can be obtained from each other by swapping adjacent members, neither of which is an ancestor of the other, one pair at a time. Suppose $\sigma'$ is obtained from $\sigma$ by swapping $V = V_{1i}$ and $V' = V_{1,i+1}$. Let $\bS = \bV_1^{<i}$, $\boldsymbol{E} = \boldsymbol{E}(\bS \cup \{V\})$, $\boldsymbol{E}' = \boldsymbol{E}(\bS \cup \{V'\})$ and $\boldsymbol{E}^* = \boldsymbol{E}(\bS \cup \{V, V'\})$. Only the factors of $V$ and $V'$ differ between $\boldsymbol{p}^\sigma(\Pr)$ and $\boldsymbol{p}^{\sigma'}(\Pr)$, and both products equal $\Pr(v, v' \mid s, e^*)$ if $\Pr(v \mid s, e) = \Pr(v \mid s, e^*)$ and $\Pr(v' \mid s, e') = \Pr(v' \mid s, e^*)$. For the first, let $\boldsymbol{A}$ be the set of observed variables that are ancestors of $\bS \cup \{V\} \cup \boldsymbol{E}^*$, including these variables. Since $\sigma$ and $\sigma'$ are consistent with a topological ordering, $V$ is not an ancestor of a member of $\bS$ or of $V'$, and therefore not of any external parent of a member of $\bS \cup \{V, V'\}$. Thus $V$ has no descendants in $\boldsymbol{A}$. By the same argument, the members of $D$ in $\boldsymbol{A}$ are $\bS \cup \{V\}$, so the district of $V$ in the subgraph on $\boldsymbol{A}$ is $\bS \cup \{V\}$, and its Markov blanket is $\bS \cup \boldsymbol{E}$. The ordered local Markov property \citep{richardson2003admgs}, which is implied by the conditional independencies, then gives $V \indep \boldsymbol{A} \setminus (\bS \cup \boldsymbol{E} \cup \{V\}) \mid \bS \cup \boldsymbol{E}$, and in particular $\Pr(v \mid s, e) = \Pr(v \mid s, e^*)$. The second identity follows in the same way. Since every closed set is $\bV_1^{\leq i}$ for some $\sigma$, this shows that $\boldsymbol{p}$ satisfies (N) and (O) for every such $\Pr$, so these equations are trivial.

Let $\bS$ be closed. The parents of a member of $\bS$ are members of $\bS$, external parents of $\bS$, or unobserved. Thus in each column of $\boldsymbol{B}$, taking the members of $\bS$ in a topological ordering shows that each of them is a function of $v_{\boldsymbol{E}(\bS)}$. The margin $p_{\bS}$ of the column is the indicator that $v_{\bS}$ equals this function, so the column satisfies (O), and clearly (N). These equations are linear, so they hold on $\mathcal{A}$, and $\mathcal{A} \cap \boldsymbol{\check{P}} \subseteq \boldsymbol{\check{P}}_O$.

\emph{Step 2: every strictly positive $\boldsymbol{q} \in \boldsymbol{\check{P}}_O$ is $\boldsymbol{p}(\Pr)$ for a strictly positive $\Pr$ that satisfies the conditional independencies and all constraints of the other districts.} 

For every district $D' \neq D$, let $\boldsymbol{p}_{D'}$ be the average of the columns of its matrix $\boldsymbol{B}_{D'}$, which is strictly positive, lies in its district polytope, and satisfies (N) and (O) for $D'$ by Step 1. Let $\boldsymbol{p}_D = \boldsymbol{q}$, and define
\[
\Pr(v) = \prod_{D'} p_{D'}(v_{D'} \mid v_{\bV_2(D')}),
\]
where the product is over all districts and $\bV_2(D')$ are the external parents of $D'$. For a closed set $\bS$ of members of $D'$, write $m_{\bS}$ for the margin of $\boldsymbol{p}_{D'}$ over the other members; by (O) it is a function of $v_{\bS}$ and $v_{\boldsymbol{E}(\bS)}$, and $m_\emptyset = 1$ by (N). Let $\boldsymbol{A}$ be a set of observed variables that contains all observed ancestors of its elements. Then each $D' \cap \boldsymbol{A}$ is closed, and we claim that
\begin{equation}
\sum_{v_{\bV \setminus \boldsymbol{A}}} \Pr(v) = \prod_{D'} m_{D' \cap \boldsymbol{A}}(v_{D' \cap \boldsymbol{A}} \mid v_{\boldsymbol{E}(D' \cap \boldsymbol{A})}). \label{eqn:ancmargin}
\end{equation}
For $\boldsymbol{A} = \bV$ this is the definition of $\Pr$. If \eqref{eqn:ancmargin} holds for $\boldsymbol{A}$ and $V \in \boldsymbol{A}$ has no children in $\boldsymbol{A}$, then $V$ appears on the right only in the factor of its own district, and summing that factor over $v_V$ gives the factor for $\boldsymbol{A} \setminus \{V\}$. Removing the variables outside $\boldsymbol{A}$ one at a time in reverse topological order proves the claim. With $\boldsymbol{A} = \emptyset$, it shows that $\Pr$ sums to one, so $\Pr$ is a strictly positive distribution. Dividing \eqref{eqn:ancmargin} for $\boldsymbol{A}$ by \eqref{eqn:ancmargin} for $\boldsymbol{A} \setminus \{V\}$ shows that $\Pr(v_V \mid v_{\boldsymbol{A} \setminus \{V\}})$ depends only on the values of $D_V \cap \boldsymbol{A}$ and $\boldsymbol{E}(D_V \cap \boldsymbol{A})$, where $D_V$ is the district of $V$. These are the district of $V$ in the subgraph on $\boldsymbol{A}$ and its parents, so $\Pr$ satisfies the ordered local Markov property, and hence all conditional independencies implied by $\mathcal{G}$ \citep{richardson2003admgs}. Finally, take $\boldsymbol{A}$ to be the variables up to $V_{1i}$ in a topological ordering whose restriction to $D$ is $\sigma_0$. The same division gives $\Pr(v_{1i} \mid \text{earlier variables}) = m_{\bV_1^{\leq i}} / m_{\bV_1^{< i}}$, which depends only on $c_i$, so it equals $\Pr(v_{1i} \mid c_i)$. The product over $i$ telescopes to $m_{\bV_1} = \boldsymbol{q}$, so $\boldsymbol{p}(\Pr) = \boldsymbol{q}$. The same argument shows that the district factor of every other district $D'$ is $\boldsymbol{p}_{D'}$, which satisfies all constraints of $D'$.

\emph{Step 3: proof of (i).} 

By Step 1, every linear combination of (N) and (O) is trivial. Conversely, suppose that $g^\top \boldsymbol{p} = d$ is trivial. By Step 2, it holds at every strictly positive point of $\boldsymbol{\check{P}}_O$. The vector with all entries equal to one over the number of values of $\bV_1$ is such a point, and the strictly positive points of $\boldsymbol{\check{P}}_O$ contain a neighbourhood of it in the affine set $L$. An affine equation that holds on a neighbourhood of a point of $L$ holds on all of $L$, and an affine equation that holds on a nonempty affine set $\{\boldsymbol{p} : M\boldsymbol{p} = a\}$ is a linear combination of the equations $M\boldsymbol{p} = a$. Thus $g^\top \boldsymbol{p} = d$ is a linear combination of (N) and (O). By the same fact, the equality constraints of $D$, which are the affine equations that hold on $\mathcal{A}$, are the linear combinations of the rows of $H_e$. By Step 1 they include (N) and (O), and the trivial ones are the linear combinations of (N) and (O), which gives the count. For the last statement, if $\Pr$ satisfies the conditional independencies and the chosen rows, then $\boldsymbol{p}(\Pr)$ satisfies (N) and (O) by Step 1, and hence every row of $H_e$.

\emph{Step 4: proof of (ii).} 

Let $F$ be the facet defined by $h^\top \boldsymbol{p} \leq c$, and $f(\boldsymbol{p}) = c - h^\top \boldsymbol{p}$. Since $\bar\beta$ is a strictly positive combination of all columns of $\boldsymbol{B}$, $f(\bar\beta) > 0$. Suppose first that $p_k = 0$ at every column of $\boldsymbol{B}$ on $F$. Since $p_k \geq 0$ on $\mathcal{P}_D$, the set $\{\boldsymbol{p} \in \mathcal{P}_D : p_k = 0\}$ is a face of $\mathcal{P}_D$. It contains $F$, and it is not all of $\mathcal{P}_D$, so it equals $F$. Hence $f$ and $p_k$ are affine functions on $\mathcal{A}$ that vanish on the affine hull of $F$, which is a hyperplane in $\mathcal{A}$, and neither vanishes on all of $\mathcal{A}$. Thus $f = \alpha p_k$ on $\mathcal{A}$ for some $\alpha \neq 0$, and $\alpha > 0$ because both are positive at $\bar\beta$. Therefore $f(\boldsymbol{p}) = \alpha p_k \geq 0$ for every $\boldsymbol{p} \in \boldsymbol{\check{P}}$ with $H_e \boldsymbol{p} = b_e$, so the constraint is trivial. Moreover, $f - \alpha p_k$ vanishes on $\mathcal{A}$, so it is a linear combination of the equations $H_e \boldsymbol{p} = b_e$.

Now suppose that no such $k$ exists, and let $\bar x$ be the average of the columns of $\boldsymbol{B}$ on $F$. Then every entry of $\bar x$ is positive. Since $\bar x$ lies in the relative interior of $F$, it satisfies every other row of $H_i$ strictly. For $\varepsilon > 0$ let $\boldsymbol{q} = (1 + \varepsilon)\bar x - \varepsilon \bar\beta$. Then $\boldsymbol{q} \in \mathcal{A}$ and $f(\boldsymbol{q}) = -\varepsilon f(\bar\beta) < 0$, and for small enough $\varepsilon$, $\boldsymbol{q}$ is strictly positive and satisfies every other row of $H_i$. Thus $\boldsymbol{q}$ lies in $\boldsymbol{\check{P}}$, satisfies the equality constraints of $D$ and violates $h^\top \boldsymbol{p} \leq c$, which is therefore nontrivial. By Step 1, $\boldsymbol{q} \in \boldsymbol{\check{P}}_O$, and Step 2 gives a strictly positive $\Pr$ that satisfies the conditional independencies and all constraints of the other districts, with $\boldsymbol{p}(\Pr) = \boldsymbol{q}$. This $\Pr$ satisfies every constraint of $D$ except $h^\top \boldsymbol{p} \leq c$.

\subsection*{Proof of Remark \ref{prop:ineqstructure}}

Let $\boldsymbol{B}$ and $\mathcal{P}_D$ be the compatibility matrix and district polytope of $D$, as in the proof of Proposition \ref{prop:nontrivial}. The equality constraints of $D$ define the affine hull of $\mathcal{P}_D$, so by Definition \ref{def:nontrivial} it suffices to find $\boldsymbol{t} \in \operatorname{aff}(\mathcal{P}_D) \cap \boldsymbol{\check{P}}$ with $\boldsymbol{t} \notin \mathcal{P}_D$, which then violates some row of $H_i$.

Let $\mathcal{G}_0$ be obtained from $\mathcal{G}$ by deleting every directed edge into a member of $D$ except those on the paths in (i) or (ii), and $A \rightarrow B$ in (i). A response function in $\mathcal{G}_0$ is a response function in $\mathcal{G}$ that ignores some parents, so the district polytope of $D$ in $\mathcal{G}_0$ is contained in $\mathcal{P}_D$ \citep[Lemma 3]{ansanelli2025observationalpartialordercausal}. Fix two levels $0, 1$ of each member and a level $e^1$ of each external parent at the start of a path. Setting the members off the paths to $0$, the other members on each path except its last to the indicator that its first vertex takes the level $e^1$, and $A$ and $B$ to the values they take in the binary instrumental variable district $Z \rightarrow A \rightarrow B$ in (i), or the Bell district $X \rightarrow A \leftarrow U \rightarrow B \leftarrow Y$ in (ii), with this indicator in place of $Z$, $X$ or $Y$, gives a linear map into the space of $\boldsymbol{p}$. It sends probability vectors to probability vectors, and columns to columns of the district polytope of $D$ in $\mathcal{G}_0$. In Section \ref{sec:nontriv}, the only equality constraints of these two districts are (N) and, in the Bell district, (O). Let $\boldsymbol{t}$ be the image of $\Pr(A = 0, B = z \mid Z = z) = 1$ in (i), or of the distribution $\Pr(A \oplus B = xy \mid X = x, Y = y) = 1$ with uniform margins \citep{popescu1994quantum} in (ii). Both satisfy these equality constraints, so $\boldsymbol{t} \in \operatorname{aff}(\mathcal{P}_D) \cap \boldsymbol{\check{P}}$.

It remains to show that $\boldsymbol{t} \notin \mathcal{P}_D$, and only this step uses the exclusion conditions. In (i), $E$ and the vertices on its path other than $A$ are not parents of $B$. So in each column of $\boldsymbol{B}$, if $A$ and the other members $\boldsymbol{T}$ that are parents of $B$ take the same values at two values of $\bV_2$ that differ only in $E$, so does $B$. Hence \eqref{eqn:ivtype} holds at every column and so on $\mathcal{P}_D$. But $\boldsymbol{t}$ has $A = 0$, $\boldsymbol{T} = 0$ and $B$ equal to the indicator of $E = e^1$, so the left side of \eqref{eqn:ivtype} with $a = 0$, $t = 0$, $b = 0$, $b' = 1$ and $E$ changed from another level to $e^1$ is $2$. In (ii), by induction along a topological ordering, a member depends in each column on $v_2$ only through the external parents with a path in $D$ to it. Thus $A$ does not depend on $E_B$, and $B$ not on $E_A$, so the CHSH inequality holds on $\mathcal{P}_D$ by the usual argument \citep{clauser1969proposed}, and $\boldsymbol{t}$ violates it.

\subsection*{Proof of Remark \ref{prop:merge}}

Let $U^*$ in $\mathcal{G}$ be the vector of the unobserved variables that were merged. Each observed variable in $\mathcal{G}$ can then depend on $U^*$ only through the components that are its parents in $\mathcal{G}'$, so every structural model for $\mathcal{G}'$ is a structural model for $\mathcal{G}$. Hence $\mathcal{M}(\mathcal{G}') \subseteq \mathcal{M}(\mathcal{G})$, and every constraint that holds on $\mathcal{M}(\mathcal{G})$ holds on $\mathcal{M}(\mathcal{G}')$.

\end{document}